\documentclass[aps,prb,preprint,superscriptaddress]{revtex4-2}

\usepackage{graphicx}  
\usepackage{color}
\usepackage{hyperref}
\usepackage{placeins}

\usepackage{amsmath}
\usepackage{amssymb}
\usepackage{graphicx,graphics}
\usepackage{latexsym,verbatim}
\usepackage{dcolumn} 
\usepackage{color}
\usepackage{cancel}

\begin{document}

\title{Anyons in Quantum Hall Interferometry}

\author{Matteo Carrega}
\affiliation{CNR-SPIN, Via Dodecaneso 33, 16146, Genova, Italy}

\author{Luca Chirolli}
\affiliation{NEST, Istituto Nanoscienze-CNR and Scuola Normale Superiore, 
Piazza San Silvestro 12, 56127 Pisa, Italy}
\affiliation{Department of Physics, University of California, Berkeley, California 94720, USA}

\author{Stefan Heun}
\affiliation{NEST, Istituto Nanoscienze-CNR and Scuola Normale Superiore, 
Piazza San Silvestro 12, 56127 Pisa, Italy}

\author{Lucia Sorba*}
\affiliation{NEST, Istituto Nanoscienze-CNR and Scuola Normale Superiore, 
Piazza San Silvestro 12, 56127 Pisa, Italy; 
*email = lucia.sorba@nano.cnr.it}

\begin{abstract}
The quantum Hall (QH) effect represents a unique playground where quantum coherence of electrons can be exploited for various applications, from metrology to quantum computation. In the fractional regime it also hosts anyons, emergent quasiparticles that are neither bosons nor fermions and possess fractional statistics. Their detection and manipulation represent key milestones in view of topologically protected quantum computation schemes. Exploiting the high degree of phase coherence, edge states in the QH regime have been investigated by designing and constructing electronic interferometers, able to reveal the coherence and statistical properties of the interfering constituents. Here, we review the two main geometries developed in the QH regime, the Mach-Zehnder and the Fabry-Perot interferometers. We present their basic working principles, fabrication methods, and the main results obtained both in the integer and fractional QH regime. We will also show how recent technological advances led to the direct experimental demonstration of fractional statistics for Laughlin quasiparticles in a Fabry-Perot interferometric setup.
\end{abstract}

\maketitle

\section*{Introduction}

Over the last forty years, quantum Hall (QH) physics has represented an extremely active research topic on the condensed matter agenda \cite{vonkl2020}. Since its discovery, the quantum Hall effect has represented a unique platform that has become a metrological realization of electrical resistance and, at the same time, displays a number of highly exotic phenomena.  More recently, the quantum Hall system has been recognized as the first known example of topological quantum matter \cite{TKNN,Haldane-NobelLecture,Stern2008,Nayak2008}.  The topological nature of QH states is at the heart of the precise and robust quantization of resistance and of the long coherence length of electrons that allow for ballistic transport over remarkable distances. These interesting features have been investigated and exploited both in the integer quantum Hall (IQH) regime and for the intriguing fractional (FQH) case \cite{Tsui1982, Laughlin1983}, where many-body effects play a crucial role in forming a strongly correlated stable system, exhibiting 
highly exotic emergent quasiparticles, that are very promising for topological quantum computation purposes \cite{Nayak2008}.
 
The topological nature of QH systems is reflected in the coexistence of an insulating bulk and metallic edge states, which are responsible for transport properties. Owing to the high degree of phase coherence of electrons, these 1D propagating channels have been viewed as the electronic analog of optical waveguides, opening an entire new field called electron quantum optics \cite{Bocquillon2014,RosenowPRL2016,Roussel2017,Glattli2017,Bauerle2018} that holds promise for quantum computation with flying qubits \cite{Ionicioiu2001,Stace2004,Feve2008,Giovannetti2008,Bordone2019,Shimizu2020}. In this setting, interferometry in the QH regime has emerged as a unique playground to study and exploit the coherence of propagating quasiparticles. Several electronic interferometer setups have been put forward and successfully realized in recent years, both constituting useful novel metrological applications and, more importantly, revealing information on Coulomb interactions and the nature of emergent quasiparticles. Since quasiparticle statistics can be inferred by moving one entity around another, electronic analogues of wave interferometers are ideal platforms to investigate these peculiar features, especially  in the FQH regime, whose excitations are predicted to be anyons \cite{Nayak2008, Stern2008, Leinaas1977, Wilczek1982}, instead of common fermions or bosons. 
 
The scope of this Technical Review is to give a survey of the main and most recent results achieved in QH interferometry. Of course, it is not possible to cover all aspects investigated in the vast literature present, where already excellent reviews have been published on different topics \cite{Stern2008,Nayak2008,Halperin2020}. Here, we describe and compare the two main geometries realized so far, namely Mach-Zehnder and Fabry-Perot electronic interferometers. We discuss their basic working principles, fabrication methods, drawbacks, and the main results obtained both in the integer and fractional QH regime. We also discuss a recent landmark experiment in Fabry-Perot interferometry that allowed direct demonstration of fractional statistics for Laughlin quasiparticles \cite{Nakamura2020}. We emphasize that another recent experiment \cite{Bartolomei2020} achieved the same milestone, employing a beam-splitter setup with noise detection. These impressive results open the way to new experiments involving more complex FQH states, with the aim of demonstrating the emergence of quasiparticles with non-Abelian statistics, a key milestone for topological quantum computation purposes \cite{Nayak2008}.

\section*{Anyons and Exchange Statistics}

Quantum statistics is at the heart of quantum mechanics, reflecting the symmetry properties of a many-body wave function under the exchange of two identical particles. In three dimensions, only two kinds of particles can exist, bosons or fermions, which are symmetric or anti-symmetric under particle exchange, respectively. In two spatial dimensions (2D), the situation turns out to be not so restricted and much richer, as realized over 40 years ago \cite{Leinaas1977} \cite{Wilczek1982} --- an adiabatic process in which two particles are exchanged twice is equivalent to a process where one particle forms a closed loop around the other particle. Indeed, in 2D, a closed trajectory encircling a particle cannot be deformed into a point without "crossing" the encircled particle and the whole system does not necessarily come back to the same state \cite{Leinaas1977, Wilczek1982, Nayak2008, Stern2008}. This idea defines the concept of non-trivial winding. After two consecutive particle interchanges (or one {  particle moved} around the other in a closed loop) the two-particle wave function, $\psi$, can acquire a non-trivial geometric phase, $\theta$,
\begin{equation}
\psi({\bf r}_1,{\bf r}_2)\to e^{i\theta/2}\psi({\bf r}_2,{\bf r}_1) \to e^{i\theta}\psi({\bf r}_1,{\bf r}_2)~,
\end{equation}
with $\theta=0$ for bosons and $\theta=2\pi$ for fermions, where ${\bf r}_1$ and ${\bf r}_2$ define the position of each particle. In 2D, other generic $\theta$ values are allowed, and quasiparticles obeying such generalized statistics are called anyons \cite{Leinaas1977,Wilczek1982,Wilczek1990,Stern2008}. 

More generally, in a system of $n$ quasiparticles (with non-degenerate ground state), an adiabatic process consisting of swapping some of the quasiparticle positions results in a phase factor, which is composed of a trivial dynamical phase and a { statistical} contribution. {  The latter can be expressed as the Berry phase \cite{Berry1984,Nayak2008} of the relative motion and} can be quantified as
\begin{equation}
\label{eq:berry}
\theta= i\oint d {\bf r} \cdot\langle \psi ({\bf r})|\nabla_{\bf r}|\psi({\bf r}) \rangle,
\end{equation}
which is independent of the microscopic details of the particle trajectories { and represents a two-particle property of the wave function}. { Besides, a single-particle Berry phase can be acquired 
separately by each particle.} A well-known example is provided by the Aharonov-Bohm (AB) phase \cite{Aharonov1959}, that is acquired by a particle of effective charge $e^*$ encircling a region threaded by a magnetic flux. In presence of a vector potential ${\bf A}({\bf r})$ describing a magnetic field of magnitude $B$, the gradient $\nabla_{\bf r}$ is replaced by $\nabla_{\bf r}-ie^* {\bf A}({\bf r})/\hbar$, and the Berry phase acquires an additional term, $\theta\to \theta+\theta_{AB}$, with 
\begin{equation}
\theta_{AB}=\frac{e^*}{\hbar}\oint d {\bf r} \cdot{\bf A}({\bf r})=2\pi (e^*/e)B \cdot A/\phi_0,
\end{equation}
where $A$ represents the area enclosed by the loop, $\phi_0=h/e$ is the flux quantum, and $e$ the electron charge. The AB phase does not depend on the details of the particular trajectory, but only on the flux threading the enclosed area. 

In the FQH regime,  the competition between the high magnetic field, the low electronic density and the Coulomb interaction, together with the two-dimensional character of the system, promotes a strongly correlated ground state wave function characterized by non-trivial exchange properties. In this regime, quasiparticle excitations are believed to be emergent anyons \cite{Wen1995, Nayak2008, Stern2008}. The simplest instance of FQH state supporting anyons is provided by the Laughlin state, that describes the series with fractional filling factor $\nu=1/(2m+1)$, with $m$ an integer. In this case, evaluation of the Berry phase leads to a statistical angle $\theta_\nu=2\pi/(2m+1)$ and an effective fractional charge $e^*=e/(2m+1)$ \cite{Laughlin1983,Arovas1984}. Indeed, the fractional quasiparticle exchange properties follow from the statistical { (two-particle)} part of the Berry phase \cite{Arovas1984,Nayak2008}, while the fractional effective charge is associated with the {  single-particle} Aharonov-Bohm  effect via the flux acquired along a closed path. 

The existence of fractional charges has been observed and confirmed with several techniques, such as the measurement of the Fano factor  in shot noise experiments \cite{dePicciotto1997,Saminadayar1997}. Detection of fractional statistics is more subtle, and after decades of intense research efforts, a first direct proof of fractional statistics in the FQH regime has been reported very recently in anyon collision experiments \cite{Bartolomei2020} and soon after confirmed also in interferometry experiments \cite{Nakamura2020}. { The former study realizes collision experiments at a beam-splitter {  based on the Hanbury-Brown-Twiss or Hong-Ou-Mandel effect} \cite{Bocquillon2014,RosenowPRL2016,Roussel2017,Glattli2017,Bauerle2018,Bartolomei2020}, that can probe the statistics of colliding particles via two-particle interference. The latter is based on interference among different paths, and several proposals that aim at probing the fractional statistics have been engineered for FQH states. Below we focus on} two of the most common electronic interferometers that have been proposed and realized in the last years for this purpose (see Box \ref{Box1} for their optical analog), and we describe the recent results obtained in this field.

\section*{Mach-Zehnder Interferometer}

\begin{figure}[h!] 
   \includegraphics[width=0.8\textwidth]{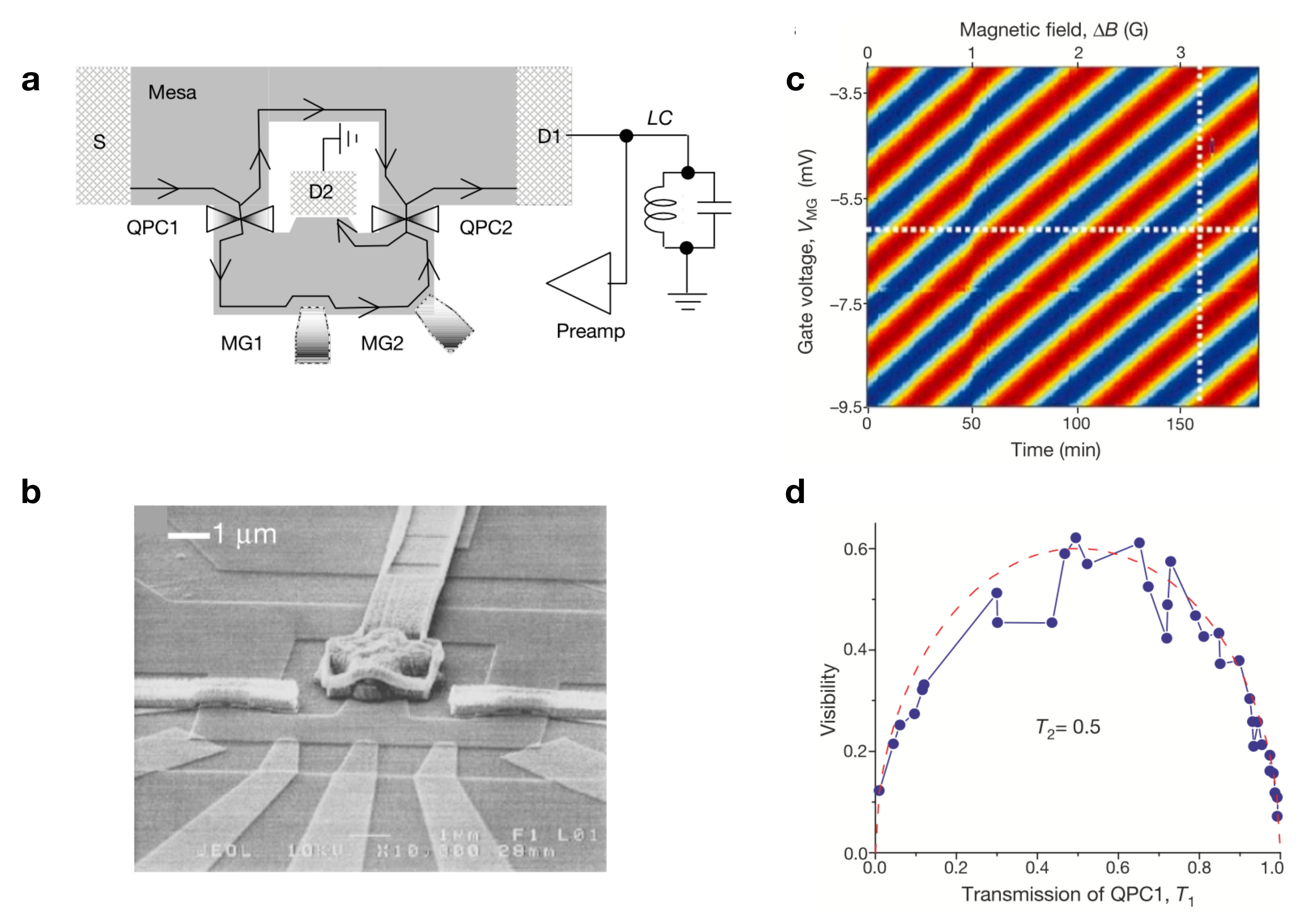}
   \caption{{\bf The electronic Mach Zehnder interferometer.}\label{Fig:MZI}(a) Scheme of the electronic Mach-Zehnder interferometer (MZI) in a Hall-bar in the Corbino geometry, featuring a hole at the center. {  For filling factor $\nu=2$, the inner edge channel (not shown) is completely backscattered at the quantum point contacts (QPC)s. The outer channel coming out from source S (represented by a black line) is partially transmitted at QPC1 and QPC2 and can therefore mix with the outer channel that flows around the hole at the center of the mesa. The outgoing signal is collected at drains D1 and D2.} Modulation gates, MG1 and MG2, allow for modulation of the area. (b) Scanning electron micrograph of an electronic MZI. A central ohmic contact, serving as a detector (D2), is connected through a metallic air bridge. QPC1 and QPC2 obtained via metallic air bridges act as beam-splitters, and five metallic gates are used as MGs. (c) Aharanov-Bohm oscillations as a function of magnetic field and modulation gate voltage, showing the typical pyjama pattern. {  Color code: maxima (red) and minima (blue).} (d) Visibility of the Aharanov-Bohm oscillation as a function of QPC1 transmission $T_1$, { together with a fit (dashed line) with the expectation $\eta\sqrt{T_1(1-T_1)}$, giving $\eta=0.62$}. All panels adapted from Ref.~\cite{Ji2003}.}
\end{figure}

\subsection*{Geometry and Basic Principles}

An electronic Mach-Zehnder interferometer can be realized in the QH regime by exploiting the edge states as electronic channels, in analogy with photon waveguides (Box~\ref{Box1}). The high degree of phase coherence in the QH regime allows for the construction of an interferometer on a scale of tens of micrometers.  Due to the chiral nature of the edge modes, dictated by the magnetic field, a MZI in the QH regime is usually realized in a Corbino geometry that features a hole at the center of the Hall-bar, as shown in Fig.~\ref{Fig:MZI}a-b. Beam-splitters are realized through quantum point contacts (QPCs) (Box \ref{Box2}) that bring the edge channels in close proximity, allowing for tunneling and (partial) backscattering. {  Depending on the structure of the edge channels that characterize an integer filling factor $\nu$,  typically only one edge channel is partially transmitted at the QPCs, and the other channels are completely reflected or transmitted.} The schematic in Fig.~\ref{Fig:MZI}a highlights {  only the scattering channels that coherently mix at the QPCs} and allows for a clear comparison with the optical analog (see Box \ref{Box1}). Propagation along the two paths introduces a relative phase, which comprises a dynamical phase $\theta_L$ and the Aharonov-Bohm phase $\theta_{AB}$. The dynamical phase depends on the energy via the momentum $k=\epsilon/(\hbar v_\epsilon)$, with $v_\epsilon$ the velocity at energy $\epsilon$. The Aharonov-Bohm phase is given by $\theta_{AB}=2\pi \phi/\phi_0$, with $\phi=B \cdot A$ the flux of the magnetic field $B$ through the area $A$ enclosed by the two paths. Interference fringes can be observed in the output currents measured at the detectors D1 and D2, that are expected to show out-of-phase sinusoidal oscillations. 

The electronic MZI was originally realized in the IQH regime in 2003 \cite{Ji2003}. The device was fabricated in a high mobility two-dimensional electron gas 
(2DEG) embedded in a GaAs-AlGaAs heterojunction, by shaping the mesa in a 3~$\mu$m wide ring (Fig.~\ref{Fig:MZI}b). An air bridge was constructed to make an Ohmic contact with the drain D2. The AB phase is modulated by changing the magnetic field $B$ or by changing the area $A$ through different modulation gates (MG) placed along one path, as shown in Fig.~\ref{Fig:MZI}a-b. The interference pattern was measured at filling factor $\nu=2$, {  with the inner edge channel completely reflected by the QPCs}. The current $I$ as a function of the modulation gate voltage and the magnetic field shows a neat interference pattern in the form of a characteristic striped color plot, the so-called ``pyjama pattern", as reported in Fig.~\ref{Fig:MZI}c.
 
\subsection*{Implementation and Results}

\begin{figure}[h!] 
   \includegraphics[width=0.8\textwidth]{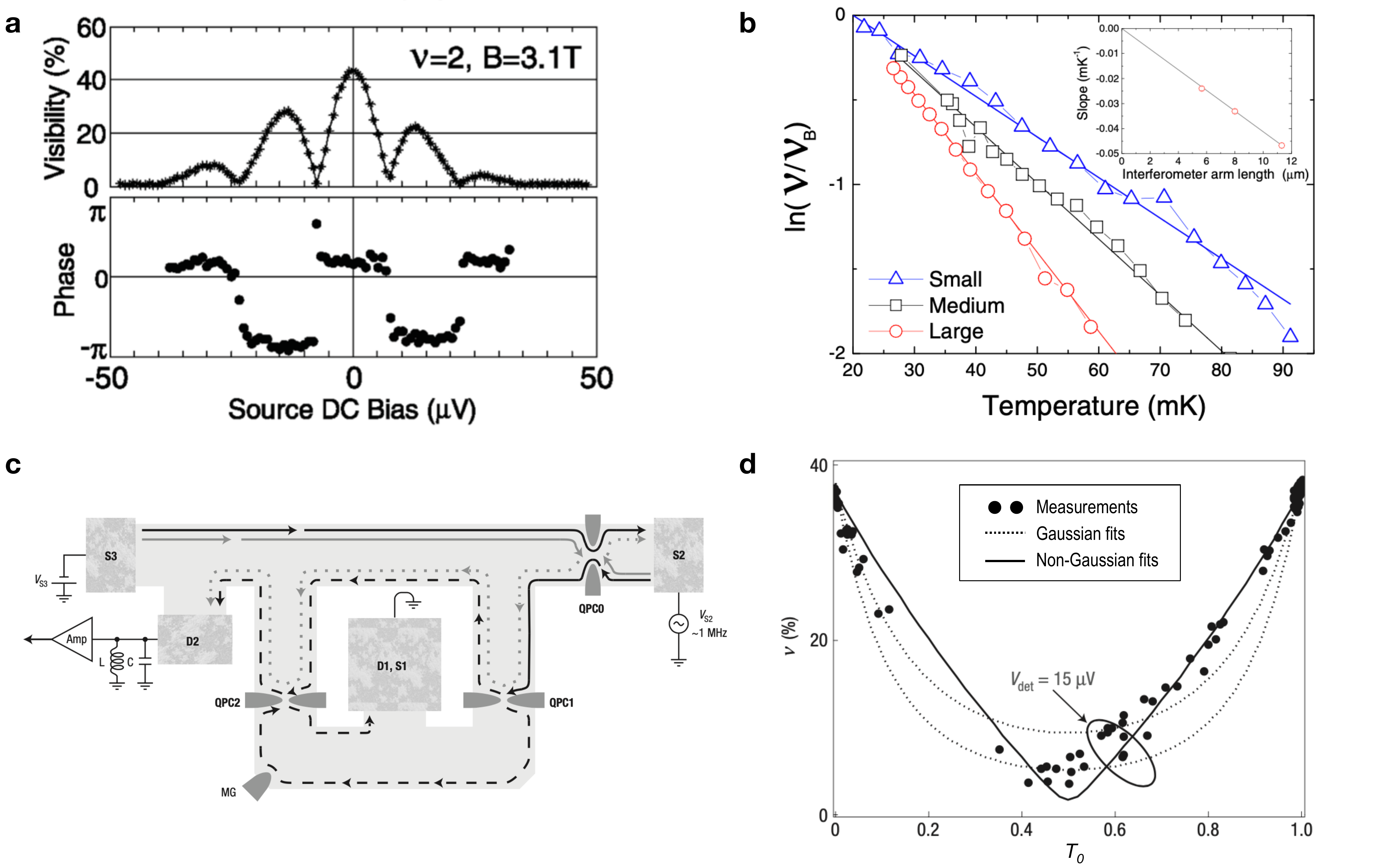}
   \caption{{\bf Visibility and coherence of Aharonov-Bohm (AB) oscillations. }\label{Fig:Lobes-MZI}(a) Visibility of the Aharanov-Bohm oscillations of a Mach-Zehnder interferometer (MZI) in the $\nu=2$ integer quantum Hall  (IQH) regime as a function of the source-drain bias, showing a lobe structure { (top panel)} and phase rigidity with $\pi$ jumps  { (bottom panel)}. (b) Log of the visibility, $\cal V$, of the AB oscillations as a function of temperature for MZIs of different size. Inset: slope of the curves versus size of the devices. { An exponential decay with temperature and size is extracted.} (c) Schematics of a MZI in the $\nu=2$ IQH regime showing the interfering outer channel (black solid/dashed line), sourced by S2, and the inner channel (gray solid/dashed line). The latter is independently biased, through contact S3, and partitioned, through quantum point contact (QPC0), to induce a controlled dephasing. Currents are collected in drains D1 and D2. The interfering area is varied through a modulation gate MG. (d) Resulting visibility as a function of the QPC0 transmission $T_0$ for $V_{\rm det}=V_{S3}$. Fits show the predictions for dephasing induced by Gaussian { (dotted lines)} and non-Gaussian  noise { (black line, see equation in the text)}. Panel a adapted from Ref.~\cite{Neder2006a}. Panel b adapted from Ref.~\cite{Roulleau2008}. Panels c and d adapted from Ref.~\cite{Neder2007c}.}
\end{figure}

The degree of coherence of the MZI is assessed by studying the visibility, $\cal V$, of the oscillations given by ${\cal V}=(I_{\rm max}-I_{\rm min})/(I_{\rm max}+I_{\rm min})$. Setting the transmission of QPC2 to $T_2=0.5$, the visibility as a function of transmission $T_1$ of QPC1 is expected to be ${\cal V}=2\eta\sqrt{T_1(1-T_1)}$. The parameter $0<\eta<1$ accounts for a reduction of the visibility. A value $\eta=0.62$ was reported in Ref.~\cite{Ji2003}, as shown in Fig.~\ref{Fig:MZI}d. Deviations from visibility with $\eta=1$ are ascribed to phase averaging and dephasing. Phase averaging typically takes place when independent electrons contributing to the detected signal acquire a slightly randomized phase, { for example due to slightly different velocities appearing in the dynamical phase}. Dephasing typically occurs due to inelastic scattering with the environment and originates mainly from Coulomb interactions. An analysis of the temperature- and bias-dependence of ${\cal V}$ shows a monotonic decay over a common energy scale, and 
a study of the shot noise points to a dominant role of phase averaging, as opposed to dephasing \cite{Marquardt2004,Chung2005,Forster2005}.

Subsequent works by different groups confirmed the functionality of the MZI in the Corbino geometry and pointed out several non-trivial aspects \cite{Neder2006a,Neder2007c,Neder2007d,Litvin2007,Roulleau2007,Litvin2008a,Litvin2008,Litvin2008a,Roulleau2008,Roulleau2008a,Roulleau2008b,Roulleau2009,Bieri2009,Litvin2010,Weisz2012}. Unexpectedly, the visibility showed a multiple lobe structure as a function of the DC source-drain bias. The measurement shows that the phase of the AB oscillations stay constant, expect for abrupt changes by $\pi$ when there are minima in the visibility (Fig.~\ref{Fig:Lobes-MZI}a) \cite{Neder2006a,Neder2007c,Neder2007d,Litvin2008,Weisz2012}. This behaviour reflects strong dephasing in the system and is not caused by phase averaging. The lobe structure appears to be non-universal, and  structures with a single side lobe have been reported, as well \cite{Roulleau2007,Roulleau2008b}. For filling factor $\nu=2$, the {  origin of} the lobe structure in the visibility has been ascribed to interchannel Coulomb interactions \cite{Levkivskyi2008,Levkivskyi2009,Rosenow2012,Helzel2015}. { At long wavelengths, chiral excitations at the edge behave as bosonic waves (plasmons), with a dispersion determined by the intrachannel Coulomb interaction \cite{Chalker2007}. Interchannel Coulomb interaction mixes the two edge plasmon modes and promotes}  a fast charge mode and a slow dipole mode as eigenmodes \cite{Levkivskyi2008,Levkivskyi2009,Rosenow2012,Helzel2015}. A tunneling event at QPC1 excites both eigenmodes, which propagate at different velocities toward QPC2. At QPC2, their recombination results in beatings, and the minima of the visibility are associated with destructive interference at an energy scale set by the velocity of the slow mode.

In order to further characterize the degree of coherence of the MZI, a measurement of the coherence length as a function of temperature and interferometer arm length was reported  \cite{Roulleau2008}, showing in both cases an exponential dependence { (Fig.~\ref{Fig:Lobes-MZI}b)}. In an attempt to isolate the relevant interfering channel, the upper arm of the MZI has been divided into disjoint regions where the inner channel forms micron-scale loops  \cite{Huynh2012}. This loop-closing strategy combined with the { freezing of} internal degrees of freedom  provide an essential isolation from the environment, with a successful increment of coherence length up to 0.1~mm \cite{Duprez2019}. 

Other implementations of an electronic MZI have also been proposed and realized besides the Corbino paradigm. In an architecture for multichannel interferometry between co-propagating edge states in the IQH regime \cite{Giovannetti2008}, two channels are mixed before and after a region in which a top gate separates them and introduces a flux sensitive area. Beam mixing of the spin-resolved edge states (at $\nu=2$) has been realized through a comb of magnetic fingers, which provides a spin-flip mechanism and the correct momentum transfer through the comb periodicity \cite{Giovannetti2008,Chirolli2012,Chirolli2013,Karmakar2015}. An alternative approach has been put forward,  based on non-equilibrium population \cite{Deviatov2011,Deviatov2012} of compressible and incompressible strips \cite{Chklovskii1992, Chklovskii1993}. In this context, details of edge equilibration \cite{Paradiso2011} and edge reconstruction \cite{Paradiso2012} have been studied with lateral resolution by scanning gate microscopy.

In graphene, a MZI has been realized by exploiting spin- and valley-polarized edge channels at the boundary of a p-n junction \cite{Wei2017,Jo2021}. {  Here, conduction and valence band electrons have opposite chirality, and for samples encapsulated in h-BN,} copropagating edge states at the two sides of the p-n junction, even with the same spin, do not equilibrate \cite{Amet2014,Zimmermann2017}. Beam mixing is achieved at the physical boundaries of the junction, where an abrupt potential variation introduces scattering among same-spin states. AB oscillations with 98\% visibility have been reported, showing strong robustness to dephasing \cite{Wei2017,Jo2021}.

\subsection*{Interaction Effects and Dephasing} 

Interferometry in an electronic MZI has proven to be demanding already at integer filling factors. { In the $\nu=1$ case,  electrostatic interactions may give rise to  dephasing and to contributions of higher harmonics  \cite{Chalker2007}.} For the $\nu=2$ case, several experiments were performed to study the reaction of the system to controlled dephasing.  If the inner channel is independently biased and partitioned by a QPC0 with transmission $T_0$ placed before QPC1 (Fig.~\ref{Fig:Lobes-MZI}c), {  the injected electrons (which are fully reflected by QPC1) act as effective detectors and dephase the interfering electrons in the outer channel via a which-path detection \cite{Neder2007c}.} {  A which-path detection is a determination of which of the two possible paths the electron takes, with consequent collapse of the wave function due to loss of indistinguishability of the interfering paths.}  The visibility as a function of $T_0$ is shown to evolve from a parabolic dependence at small detector bias $V_{\rm det}$ to a V-shaped dependence at higher detector bias (Fig.~\ref{Fig:Lobes-MZI}d). { The presence or absence (with probability $T_0$ and $1-T_0$, respectively) of one detector electron gives an extra phase $\delta\theta\propto V_{\rm det}$ to the interfering electrons, yielding visibility ${\cal V}\propto \left| T_0+(1-T_0)e^{i\delta\theta} \right|$ that provides a good fit to the data (Fig.~\ref{Fig:Lobes-MZI}d), suggesting an evolution from  Gaussian noise at small bias to non-Gaussian noise at higher bias \cite{Neder2007c,Roulleau2008a}.} { A qualitative change in the lobe structure of the visibility as a function of $T_0$  was predicted to occur in the presence of strongly non-Gaussian noise, for which a structure with one lobe at low $T_0<1/2$ evolves to a structure with multiple lobes for $T_0>1/2$ \cite{Levkivskyi2009},  yielding a  noise-induced phase transition, that was recently confirmed experimentally \cite{Helzel2015}.} 

The use of Ohmic contacts embedded in one arm of the MZI has enabled an alternative characterization of the coherence of the system. When an Ohmic contact acts as a voltage probe, loss of coherence results from a which-path detection \cite{Roulleau2009}. At the same time, the charging energy of a micrometer-sized Ohmic contact may prevent full suppression of coherence for low bias and temperature \cite{Idrisov2018}, as no phase space is left for inelastic scattering, with the visibility saturating at its maximum value, which has been experimentally confirmed  \cite{Duprez2019transmitting}. 

\begin{figure}[h!] 
   \includegraphics[width=0.8\textwidth]{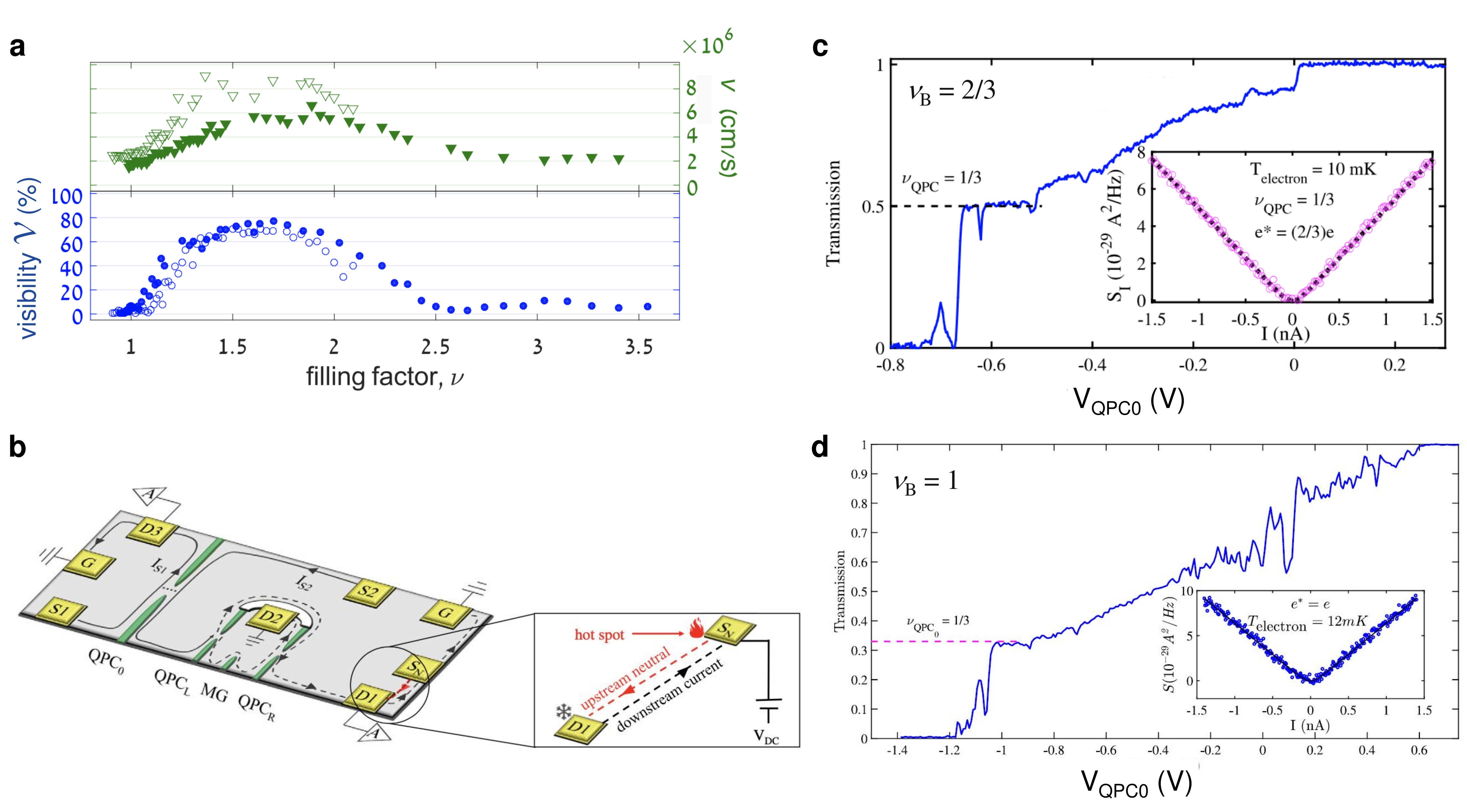}
   \caption{{\bf Impact of neutral modes on the interference signal.} \label{Fig:NeutralModes}(a) Visibility of the Aharanov-Bohm oscillations as a function of  filling factor $\nu$. Upper panel, velocity $v$ extracted from the first minimum of the visibility lobe at bias $V_0$ (cf.~Fig.~\ref{Fig:Lobes-MZI}a) via $eV_0=hv/2L$. { Lower panel, amplitude of the visibility at zero-bias. The two quantities show the same behavior with filling factor $\nu$. Full and open symbols refer to different devices.} (b) Schematics of the Mach-Zehnder interferometer obtained in a Hall-bar in the Corbino geometry, that features a hole in the center of the mesa. Quantum point contacts (QPCs) are used to mix the beams and the interfering area is varied through modulation gates (MGs). Carriers are injected via sources (S's) and collected in drains (D's). Grounded contacts are labeled as G's. Zoom in of the part of the devices allowing for independent assessment of the neutral modes propagating from source $S_N$ to drain D1. (c-d) Transmission as a function of the voltage QPC0 featuring a plateau at $\nu_{\rm QPC}=1/3$ in the case of  bulk filling factor (c) $\nu_B=2/3$ and (d) $\nu_B=1$. Insets: shot noise versus applied current showing a quasiparticle charge of (b) $e^*=(2/3)e$  and (c) $e^*=e$. Panel a adapted from Ref.~\cite{Gurman2016}. Panels b-d adapted from Ref.~\cite{Bhattacharyya2019}.}
\end{figure}

{ The evolution of visibility as a function of magnetic field has emphasized the appearance of three energy scales, associated with the temperature and bias dependence of the visibility. These scales are the inverse coherence length, the distance between the visibility minima between lobes, and the Gaussian decay scale of the visibility envelope with bias. All three scales change in a similar way with filling factor \cite{Litvin2008}. Recent measurements of} the evolution of the visibility from $\nu=3.5$ to $\nu=1$ show a maximum at $\nu=1.5$, a minimum at $\nu=2.5$, and a reduction to zero when approaching $\nu=1$ (Fig.~\ref{Fig:NeutralModes}a) \cite{Gurman2016}. The first minimum in the lobe structure of the visibility (at bias voltage $V_0$) allows us to extract a characteristic velocity $v$ as $eV_0=hv/2L$, that may be associated with a slow dipole mode. A clear correlation appears between the velocity $v$ and the visibility, with a fading of oscillations as the filling factor approaches $\nu=1$ \cite{Bhattacharyya2019}. The precise mechanism responsible for the loss of coherence at filling factor $\nu=1$ still remains an open issue.

\subsection*{Fractional Case}

An anyonic MZI for quasiparticles in the FQH regime can be designed in a similar way to that for the IQH regime, which exploits edge modes in a Corbino geometry. {  The expected interference signal depends on the structure of the edge, and the simplest case is provided by Laughlin fractions $\nu=1/(2m+1)$.  For weakly closed QPCs (Box \ref{Box2}), edge channels propagating on opposite sides of the Hall bar get closer and quasiparticles tunnel from one edge to the other traversing the Hall bulk.}  Let us assume that $n_q$ quasiparticles are present in the bulk of the MZI. Each time a quasiparticle tunnels  from one edge to the other at a QPC, the value of $n_q$ changes by one. At zero temperature, the flow is unidirectional and determined by the chemical potential difference between the two edges. At a given stage, after $j$ quasiparticles have tunneled, the rate for transition from a state with $j$ quasiparticles to a state with $j+1$ quasiparticles depends on $j~ {\rm mod} (1/\nu)$. 
{ In the limit of weak tunneling at the QPCs, the rate of quasiparticle tunneling, ${\tau_j}^{-1}$,  through the MZI   can be written as \cite{Law2006,Feldman2007}
\begin{equation}\label{Eq:RateMZIanyons}
\frac{1}{\tau_j}=c_1(|\Gamma_1|^2+|\Gamma_2|^2)+c_2[\Gamma^*_1\Gamma_2\exp\left(2\pi i\nu\phi/\phi_0+2\pi i\nu (n_q+j)\right)+c.c.],
\end{equation}
where $\Gamma_1$ and $\Gamma_2$ are the tunneling amplitudes at the first and second QPC, and $c_1$ and $c_2$ are coefficients that depend on temperature and bias voltage. A current $I=e/\tau$ is obtained when one electron is collected at the detector, after time $\tau=\sum_{j=1}^{1/\nu}\tau_j$ in which a number $1/\nu$ of quasiparticles have  tunneled \cite{Law2006,Feldman2007}. Assuming $\Gamma_2\ll\Gamma_1$, to first order in $\Gamma_2$, destructive interference takes place in the current among the $1/\nu$ different rates, and the leading flux-dependent contribution scales as
\begin{equation}
I_\Phi\sim \Gamma_2^{1/\nu}\exp(2\pi i\phi/\phi_0)+{\rm c.c.},
\end{equation} 
so that the AB oscillations in devices in the Corbino geometry show a periodicity with $\phi_0$, in agreement with the Byers-Yang theorem \cite{Byers1961,Feldman2006}.}

In addition, signatures of fractional statistics can be extracted from the Fano factor, $F=S/(2eI)$. { Here, $S$ is the current noise that is sensitive to charge fluctuations and carries information of the discrete nature of the quasiparticles involved in the tunneling process. } In a non-interferometric setup { (single QPC)} the Fano factor is typically used to extract information on the quasiparticle effective charge. In the MZI, the Fano factor for electrons for which $\nu=1$, is flux independent despite a flux dependence of current and noise. For fractional quasiparticles, a flux-dependent Fano factor is expected  \cite{Jonckheere2005,Feldman2007}.

To date, { despite tremendous experimental efforts}, AB oscillations in a MZI in Corbino geometry have never been reported for the FQH regime. { The principal reason for that is believed to be dephasing due to the appearance of neutral upstream modes that may act as a which-path detector and destroy the indistinguishability of the interfering paths}. {  Neutral modes are predicted to occur in hole-conjugate states that feature fundamental counter-propagating edge modes (see Box \ref{Box3}). Due to the Coulomb interaction, the confining potential and disorder, edge reconstruction is predicted to occur, resulting in downstream charged modes and upstream neutral modes \cite{Kane1994,Kane1995}.} In the prototypical example of $\nu=2/3$  (see Box ref{Box3}), a downstream $\nu=1$ mode is accompanied by an upstream $\nu=1/3$ mode \cite{MacDonald1990,Wen1990,Wen1990a}. Analogously to the discussed $\nu=2$ case, the interchannel Coulomb interaction promotes charge and (neutral) dipole eigenmodes, the latter propagating upstream with respect to the direction dictated by the magnetic field \cite{Kane1994,Kane1995}. {  As a consequence, quasiparticles are no longer fundamental edge excitations, and tunneling at a QPC excites both charge and neutral modes. In the MZI schematically depicted in Fig.~\ref{Fig:NeutralModes}b, a quasiparticle originating from S2 and collected at D2 may either tunnel at QPC1 and follow the path around the hole or keep propagating along the mesa edge and tunnel at QPC2. The two events may excite distinct  jets of upstream neutral mode wave packets, which are spatially separated and time delayed, as it takes the quasiparticle some time to propagate from QPC1 to QPC2. If the separation is large enough so that the neutral mode wave packets do not overlap, this excitation of neutral jets breaks the indistinguishability of the interfering paths and introduces dephasing via which-path detection. Their overlap decreases with the overall size of the MZI and can be thus recovered only by reducing the size of the system, the working temperature, or the bias voltage, posing quite severe restriction in practice \cite{Goldstein2016}.}

{ In addition to the composite nature of edge modes, an unexpected appearance of a $\nu_{\rm QPC}=1/3$ conductance plateau in the QPCs accompanied by shot noise has been reported, which suggests the presence of additional neutral modes \cite{Gurman2016,Bhattacharyya2019}}. For the device sketched in Fig.~\ref{Fig:NeutralModes}b, this is shown in Fig.~\ref{Fig:NeutralModes}c-d both for the case of bulk integer and fractional filling \cite{Bhattacharyya2019}. A conductance plateau typically indicates full transmission (or full reflection) of certain edge modes. Since no partitioning takes place, the transmitted current is expected to be noiseless. An anomalous ``noise on plateau'' was identified by independent assessment to result from excitation of upstream neutral modes at the QPC.

{ The reported phenomenology can be set in a more general framework of edge reconstruction and renormalization among multiple channels. In hole-conjugate fractions, a more complex structure featuring additional counter-propagating channels may arise possibly due to a smooth confining potential \cite{Wang2013}. Edge reconstruction and equilibration over long distances wash away the detailed structure and result in a conductance signal featuring no noise on the QPC plateaus. When probed on shorter length scales, equilibration may not be effective, and the complex edge structure may result in noisy conductance plateaus  \cite{Sabo2017}. In general, currents and noise may be affected by the presence or absence of symmetries among the edge channels \cite{Park2021}.}

Possible signatures of AB oscillations in a MZI in the FQH regime have been reported for the $\nu=1/3$ case \cite{Deviatov2012}. The geometry of this device does not contain an etched region inside the interference loop. Aharonov-Bohm oscillations with $\phi^*=(e/e^*)\phi_0=\phi_0/\nu$ have been reported {  and were ascribed to interference of fractional charge quasiparticles.}

\section*{Fabry Perot Interferometer}
\label{Sec:FPI}

\subsection*{Geometry and Basic Principles}

\begin{figure}[h] 
   \includegraphics[width=0.7\textwidth]{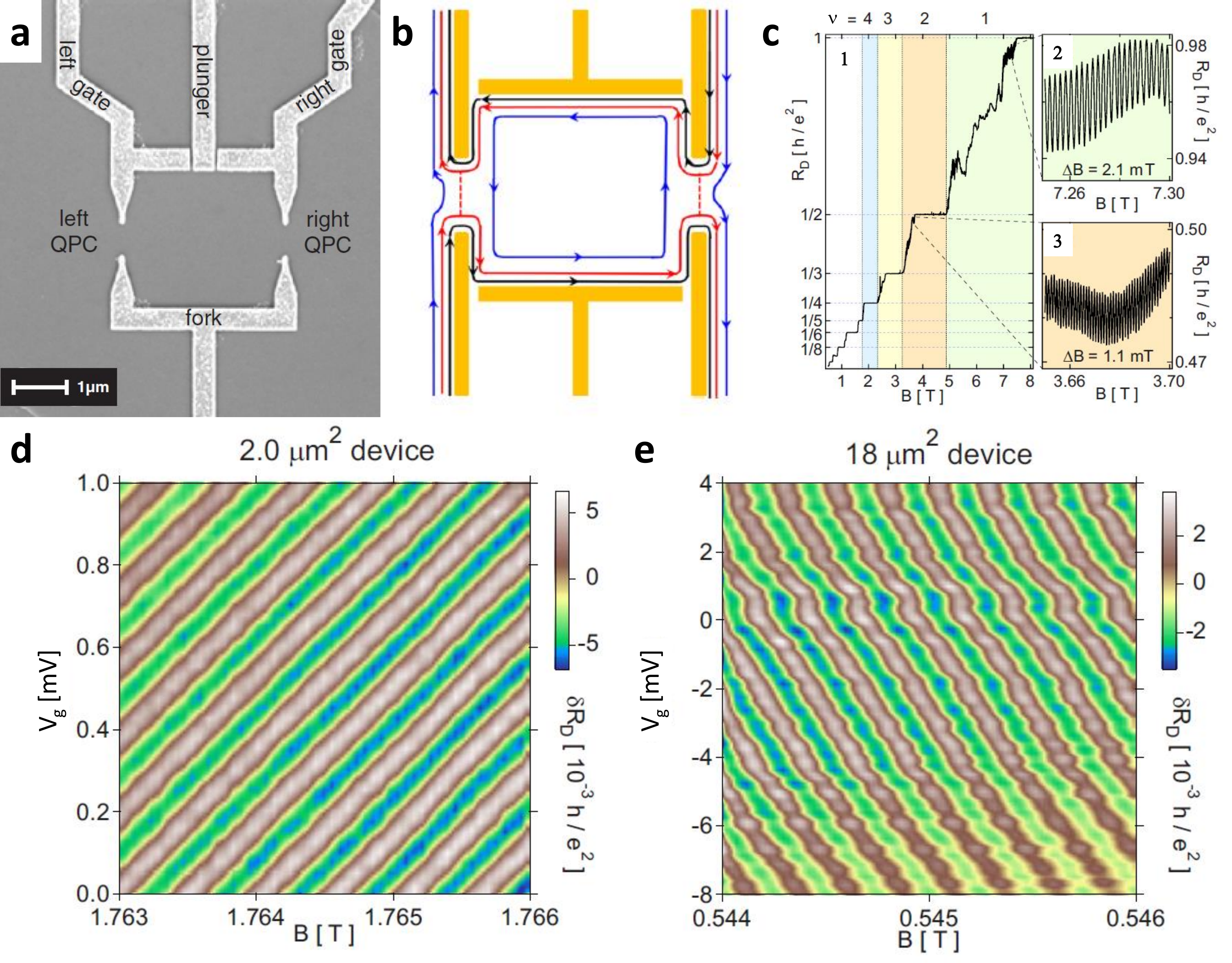}
   \caption{{\bf The electronic Fabry-Perot interferometer.} \label{fig:fpi1} (a) Representative SEM image of a Fabry-Perot interferometer (FPI) where the bright regions are the metallic gates.  (b) Schematic showing an interference path with multiple edge states in which the outermost mode is fully transmitted {($f_t=1$)}, the innermost mode is fully backscattered, and the middle mode is partially transmitted by both quantum point contacts (QPCs); in this configuration only the middle mode is interfered. (c) Oscillations in diagonal resistance, $R_D$, as a function of magnetic field $B$ for a FPI device. Panel 1 shows $R_D$ as a function of $B$, with well-quantized integer quantum Hall plateaus. Different colored backgrounds indicate different numbers of fully occupied Landau levels ($\nu$) through the device. Panels 2 and 3 report zoom--ins of the data in panel 1, at filling factors 1 and 2, respectively, showing oscillations in $R_D$, and their $B$ periods $\Delta B$. (d) $\delta R_D$, i.e.~$R_D$ with a smooth background subtracted, as a function of $B$ and modulation--gate voltage $V_g$, for a 2.0~$\mu$m$^2$ device. This plot shows a clear Coulomb-dominated behavior. (e) Same as in (d), but for a 18~$\mu$m$^2$ device. This plot shows a clear Aharonov-Bohm behavior. 
   Panel a adapted from Ref.~\cite{Ofek2010}. Panel b adapted from Ref.~\cite{Nakamura2019}. Panel c, d and e from Ref.~\cite{Zhang2009}.}
\end{figure}

A complementary geometry is the so-called electronic Fabry-Perot interferometer (FPI) \cite{Chamon1997}. Two constrictions are patterned on a Hall-bar to define two QPCs placed at a distance $d$, which play the electronic analog of two semi-transparent mirrors (Fig.~\ref{fig:fpi1}a). On a QH plateau, current carrying states counter-propagate on the two opposite edges of the Hall-bar (kept at different chemical potentials) with a quantized two-terminal Hall conductance $G_H=\nu\frac{e^2}{h}$. Interedge tunneling is {induced} by the two QPCs, which leads to a finite backscattered current (Fig.~\ref{fig:fpi1}b). Backscattering amplitudes $t_1$, $t_2$ at the QPCs, experimentally controlled by two split-gates, produce interference patterns in the backscattered current {  $I\sim |t_1+t_2 e^{i\theta}|^2$ } in the same way as partial waves interfere in the famous two slit experiment. Interference fringes depend on three main parameters: the external magnetic field $B$, {the effective area $A$ of the FPI, defined by the two edges and the two QPCs regions,} and the number $n_q$ of quasiparticles localized inside it. Quasiparticles backscattered at the constrictions will move around localized {quasiparticles} inside the FPI, thus realizing the braiding gedanken experiment in a quite natural fashion~\cite{Chamon1997, Stern2008}. Interference phase shifts are expected by changing the number $n_q$ {of quasiparticles inside the interferometer}, constituting a direct hallmark of anyonic statistics. For small tunneling amplitudes $t_i$, the backscattering current response has an oscillatory behavior, where the phase $\theta$ is a combination of the phase acquired by a quasiparticle encircling an Aharonov-Bohm flux, scaled with the effective {quasiparticle} charge $e^*$, and the anyonic statistical angle $\theta_\nu$~\cite{Halperin2011}:
\begin{equation}
\label{theta}
\theta =2\pi\frac{e^*}{e}\frac{B \cdot A}{\phi_0} + n_q \theta_\nu~.
\end{equation}
For the Laughlin sequence, the statistical angle is predicted to be {$\theta_\nu = 2\pi/(2 m+ 1)$, with $m$ an integer}. Equation (\ref{theta}) holds also for other FQH states with multiple internal edge structures~\cite{Jain1989, Inoue2014, Grivnin2014} or more exotic filling factors like 
$\nu=5/2$~\cite{Nayak2008, Stern2008}, with the proper statistical angle $\theta_\nu$ associated to the underlying quasiparticles. Abrupt changes in the phase $\theta$ upon variation of $n_q$ inside the interferometer are thus a 
consequence of the anyonic statistics. The total current backscattered by 
the interferometer will depend on $\cos(\theta)$. Therefore, the interference phase can be probed by measuring the two-terminal conductance across 
the device. In the strong backscattering regime with almost closed QPCs, the FPI resembles a quantum dot with a well-defined number of excitations. Here, rather than a sinusoidal behaviour with magnetic field or interferometer area, one expects resonance lines to appear in correspondence to degeneracy points of addition energies (Coulomb blockade) \cite{Rosenow2007, Halperin2011, NgoDinh2012}. It is worth stressing that the predicted (and also observed) periodicities do not depend on transmission amplitudes.

\subsection*{Implementation and Results}

Owing to this physical picture, great experimental efforts have been made to realize FPIs both in the integer and fractional QH regime~\cite{Wees1989, Camino2005, Camino2007, Godfrey2007, Deviatov2008, Zhang2009, Willett2009, Lin2009, Lin2009a, McClure2009, Ofek2010, McClure2012, Choi2015, Sivan2016, Sivan2018, Nakamura2019, Roosli2020, Nakamura2020}, in order to reveal non-trivial anyonic statistics. {  Early realizations of FPIs, effectively realizing dots and anti-dots in the fractional quantum Hall phases \cite{Simmons1989,Goldman1995,Maasilta2000,Goldman2001,Goldman2003}, reported oscillations  that turned out to be manifestations of fractional charge \cite{Kivelson1990,Lee1990,Thouless1991,Gefen1993}.} A typical experimental setup involves the simultaneous measurement of quantized Hall resistance $R_H$ and the diagonal resistance of the FPI $R_D$. The latter is proportional to the two-terminal Hall conductance and hence to the backscattered current signal, in a four-terminal configuration, performed at low temperature (of the order of tens of mK). Interference patterns can be inspected by varying the magnetic field $B$ or  by applying a voltage $V_g$ on a plunger  (modulation) gate which effectively shrinks and modifies the  area $A(B,V_g)$ (Fig.~\ref{fig:fpi1}c). Unexpectedly, by analyzing the obtained periodicities as a function of these external knobs, a richer panorama, compared to the naive picture exposed above, has been observed, even at integer filling $\nu$ where quasiparticles are electrons with trivial statistical angle $\theta_\nu$~\cite{Camino2005, Camino2007, Zhang2009, Willett2009, Lin2009}. It turned out that Coulomb interactions play a critical role, determining different operating regimes of FPIs. In a first generation of devices, two main regimes have been identified, namely a Coulomb-dominated and an Aharonov-Bohm dominated regime~\cite{Rosenow2007, Zhang2009, Ofek2010, Halperin2011}.

\subsection*{Coulomb Blockade limits Visibility}

Distinct behaviors have been reported analyzing different geometrical FPI sizes. A seminal work \cite{Wees1989} demonstrated resistance oscillations as a function of magnetic field in an electronic FPI, opening the way to a large amount of experiments which showed rich physics even in the (naively thought simple) case of the IQH regime~\cite{Camino2005, Camino2007, Godfrey2007, Deviatov2008, Zhang2009, Willett2009, Lin2009, Lin2009a, McClure2009, Ofek2010, McClure2012, Choi2015, Sivan2016, Sivan2018}. 

Indeed, for a small area, of the order of a few $\mu$m$^2$, strong bulk-edge capacitive coupling characterizes the FPI operation, and a change in the bulk density determines a change of the area enclosed by the edge. Therefore, in this Coulomb-dominated regime, changing $n_q$ amounts to  a continuous variation of the area of the interference loop and the phase jump becomes an unobservable integer multiple of $2\pi$ both for integer and fractional QH~\cite{Rosenow2007, Halperin2011}. Conversely, in the extreme AB regime, where the bulk and the interfering edge are not coupled, the area of the interference loop does not vary with $B$, or by changing $n_q$~\cite{Ofek2010, Sivan2016}. Therefore, it is only in this latter regime that anyonic statistics can be probed~\cite{Halperin2011}. Measurements of $R_D$ revealed different oscillating behaviors as a function of both magnetic field and side gate $V_g$. Looking at color maps of lines of constant phase in a $B-V_g$ plane, the Coulomb-dominated regime is characterized by a pyjama pattern with lines of positive slope (an increase in $B$ {is accompanied by an increase} in $V_g$, in order to keep the phase constant~\cite{Zhang2009, Ofek2010, Sivan2016} (Fig.~\ref{fig:fpi1}d). 

Similar interference fringes have been also reported considering several fractional filling factors~\cite{Camino2007, Camino2008, Zhang2009, Willett2009, McClure2012}. Importantly, it has been recognized that, together with the interfering edge, the number of fully transmitted channels $f_t$, belonging to the lowest Landau levels, comes into play in determining the periodicity of the FPI oscillations \cite{Rosenow2007, Zhang2009, Halperin2011}. Devices with front gates deposited on etch trenches allowed a fine and independent control of the filling factor, the number of edge channels in the constrictions, and the FPI area with respect to the bulk filling factor~\cite{Camino2007}. In the Coulomb-dominated regime, magnetic field periodicities turn out to be $\Delta B =\phi_0/(A f_t)$ (no oscillations when $f_t=0$); similarly, gate voltage periodicity $\Delta V_g$ depends solely on $f_t$. These periodicities can be understood assuming that the relevant area of the interferometer changes as the magnetic field is varied, with a threaded flux $\phi = B \cdot A \left( B, V_g \right)$. Therefore, the phase $\theta$ responds to small variations in magnetic field or gate voltage as
\begin{eqnarray}
\nonumber d\theta&\propto&\frac{\partial(A\cdot B)}{\partial B}dB+\frac{\partial(A\cdot B)}{\partial V_g}dV_g\\
&=&\left(A+B\frac{\partial A}{\partial B}\right)dB+B\frac{\partial A}{\partial V_g}dV_g,
\end{eqnarray}
For the number of (outer) fully transmitted channels $f_t=0$, increasing $B$ is followed by a decrease in the area $A$ in such a way that the flux, and hence the number of occupied states, is kept constant, but the charge density at the center of the FPI area grows.  A spatial imbalance between electrons and ionized donors takes place. Increasing the magnetic field further must eventually lead to a relaxation of that imbalance by {expulsion of an electron (creation of a hole)} within the interfering Landau level. The area thus "breathes" with magnetic field~\cite{Sivan2016}, and it decreases monotonically and increases abruptly while keeping {the average area} constant. The same happens for $f_t>0$, only with a faster decrease with magnetic field related to the possible transfer of charges from the interfering edge to the lower Landau levels.

Conversely, larger FPI devices (of the order of $\sim 20$~$\mu$m$^2$) showed an opposite behaviour with lines of constant phase with a negative slope~\cite{Zhang2009, Ofek2010}, consistent with the AB regime (Fig.~\ref{fig:fpi1}e)~\cite{Sivan2016}. Here, reported periodicities are no longer independent of magnetic field, and in particular $\Delta V_g \propto 1/B$, while $\Delta B\propto 1/A$. Although coherent Aharonov-Bohm and Coulomb-dominated regimes seem to be of different nature, the former being interference of independent quasiparticles and the latter being associated with periodic charging (and discharging) of the FPI with single electrons, they reflect two {sides} of the same coin, that can occur on the same device~\cite{Halperin2011}.

A unified picture for both Coulomb-dominated and Aharonov-Bohm regimes based on energetic considerations has been developed in Ref.~\cite{Halperin2011}. Physically, this model consists of a capacitance network, taking into account self-capacitance of the interfering edge, the capacitance of the localized quasiparticles, the capacitance between the bulk quasiparticles and the edge, and the capacitive coupling of the gate to the interferometer. By minimizing the total energy of the system, one can deduce  the expected periodicities of the interfering fringes, recovering the Coulomb-dominated regime with a strong capacitive bulk-edge coupling and the opposite AB regime, where this mutual capacitance is negligible.

\begin{figure}[h] 
   \includegraphics[width=0.8\textwidth]{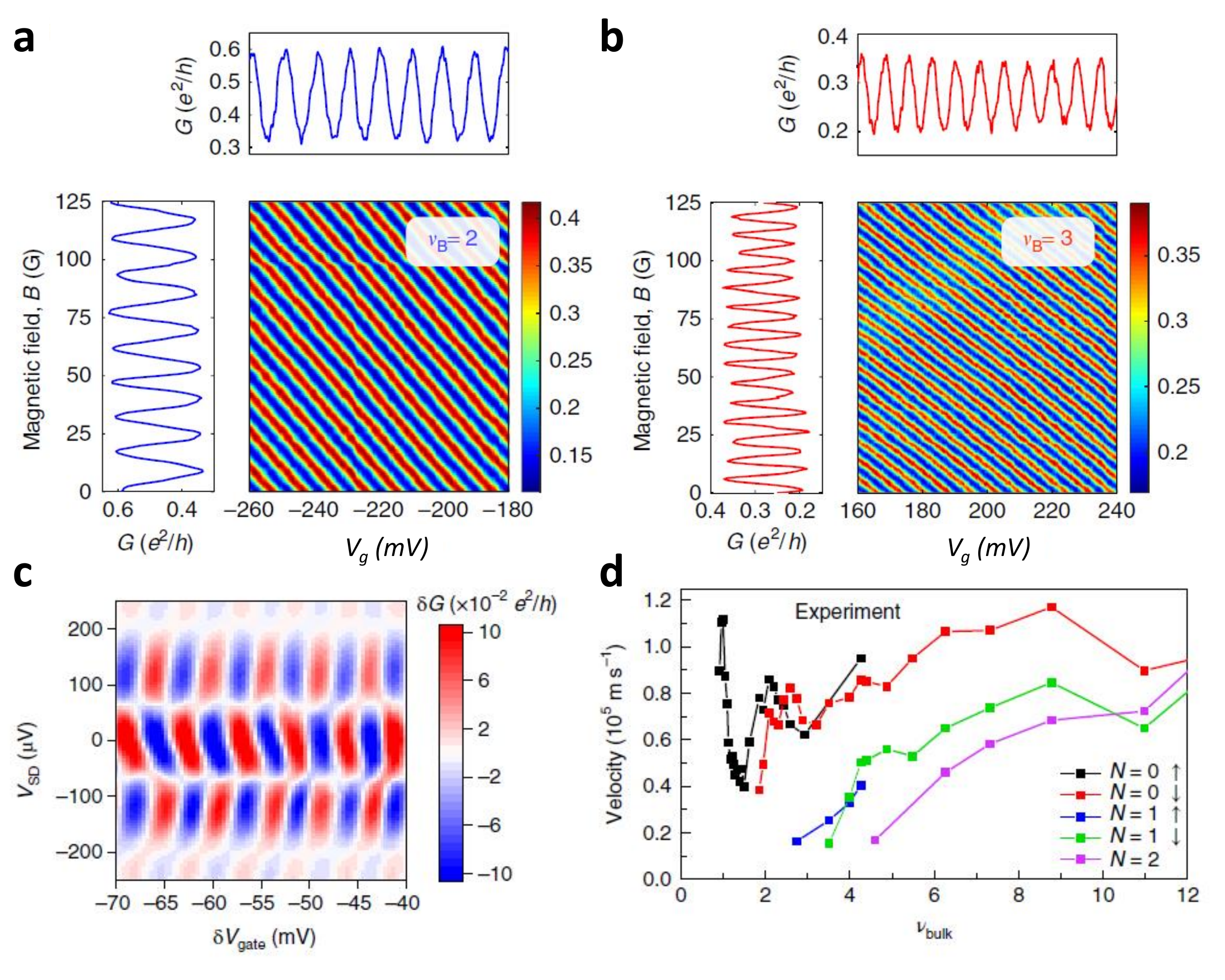}
   \caption{{\bf Coulomb-dominated versus Aharonov-Bohm regime.} \label{fig:fpi2} (a, b) Aharonov-Bohm interference in the $h/e$ and $h/2e$ regimes measured with a 2~$\mu$m$^2$ Fabry-Perot interferometer (FPI) with grounded center-Ohmic contact. Conductance $G = 1/R_D$ of the FPI versus both magnetic field $B$ and modulation--gate voltage $V_{g}$ (a) in the $h/e$ regime and (b) in the $h/2e$ regime, measured at $B = 5.2$~T (bulk filling factor $\nu_B \sim 2$) and $B = 3$~T ($\nu_B \sim 3$), respectively. (c) Differential conductance at $\nu_{bulk} = 1$ interfering the $\nu = 1$ mode as a function of side-gate voltage $\delta V_{\rm gate}$ (relative to $-1.4$~V) and source-drain voltage $V_{\rm SD}$. (d) Edge-state velocities extracted from the differential conductance oscillations for different edge modes for the $N = 0$, 1, and 2 Landau levels. Panels a and b adapted from Ref.~\cite{Choi2015}. Panels c and d adapted from Ref.~\cite{Nakamura2019}.}
\end{figure}

\subsection*{Screening unwanted Coulomb Interactions}
 
Several technological advances have been made in order to suppress electron--electron interactions and bulk-edge coupling, achieving the AB regime in a controlled fashion, even in rather small FPIs. A central top gate covering the whole interferometer has been employed to reduce the charging energy \cite{Zhang2009}; alternatively, low temperature illumination has been used to enhance screening by the doping layer \cite{Willett2009}. A more effective screening has been obtained using a grounded Ohmic contact of the size of $\sim 0.5$~$\mu$m$^2$ inside the incompressible bulk of the FPI~\cite{Ofek2010, Choi2015, Sivan2016}. Placing such an Ohmic contact in the vicinity of the FPI, instead of placing it within the FPI center, results in slightly less effective screening, allowing to inspect also the intermediate regime between AB and Coulomb. This leads to the observation of a lattice structure (checkerboard pattern) in the $B-V_g$ plane with more than one characteristic frequency \cite{Sivan2016}. However, these methods lower the charging energy of the FPI as a whole, suppressing Coulomb-dominated features, without much effect on the interchannel capacitance. Residual Coulomb interactions between different edge channels can give rise to interesting behavior even for devices operating at IQH filling in the AB regime. 

Measurements in the AB regime for FPIs of different sizes ($\sim 2-12$~$\mu$m$^2$) with a center Ohmic contact have been reported, for integer filling in the range $\nu=2$ to $\nu=4$ ~\cite{Choi2015, Sivan2018}. Monitoring the flux dependence of the conductance at $\nu>2$, an unexpected halving of the oscillation periods both in $B$ field and $V_g$, without any sign of the fundamental periodicity $h/e$ was found (Fig.~\ref{fig:fpi2}a-b) \cite{Choi2015}. To explain this unexpected doubling of AB frequencies, different scenarios have been tested. For instance, the possibility that only an even number of interfering windings can occur, has been excluded by comparing the $h/(2e)$ observed visibility and the one estimated based on the reported dephasing length~\cite{Choi2015}. The doubling of frequence has been attributed to strong interedge coupling, with an effective electron pairing among them that is consistent with quantum shot noise measurements reporting an effective charge $e^*=2e$~\cite{Choi2015, Sivan2018}. These intriguing results, reminiscent of effective attractive (Cooper-like) pairing among electrons, have been confirmed in subsequent experiments~\cite{Sivan2018, Nakamura2019} and explained~\cite{Ferraro2017, Sivan2018, Frigeri2019, Frigeri2020} in terms of mutual edge capacitances and the emergence of neutral edge modes whose exchange can mediate such counterintuitive effective attractive interactions.

\begin{figure}[h!] 
   \includegraphics[width=0.8\textwidth]{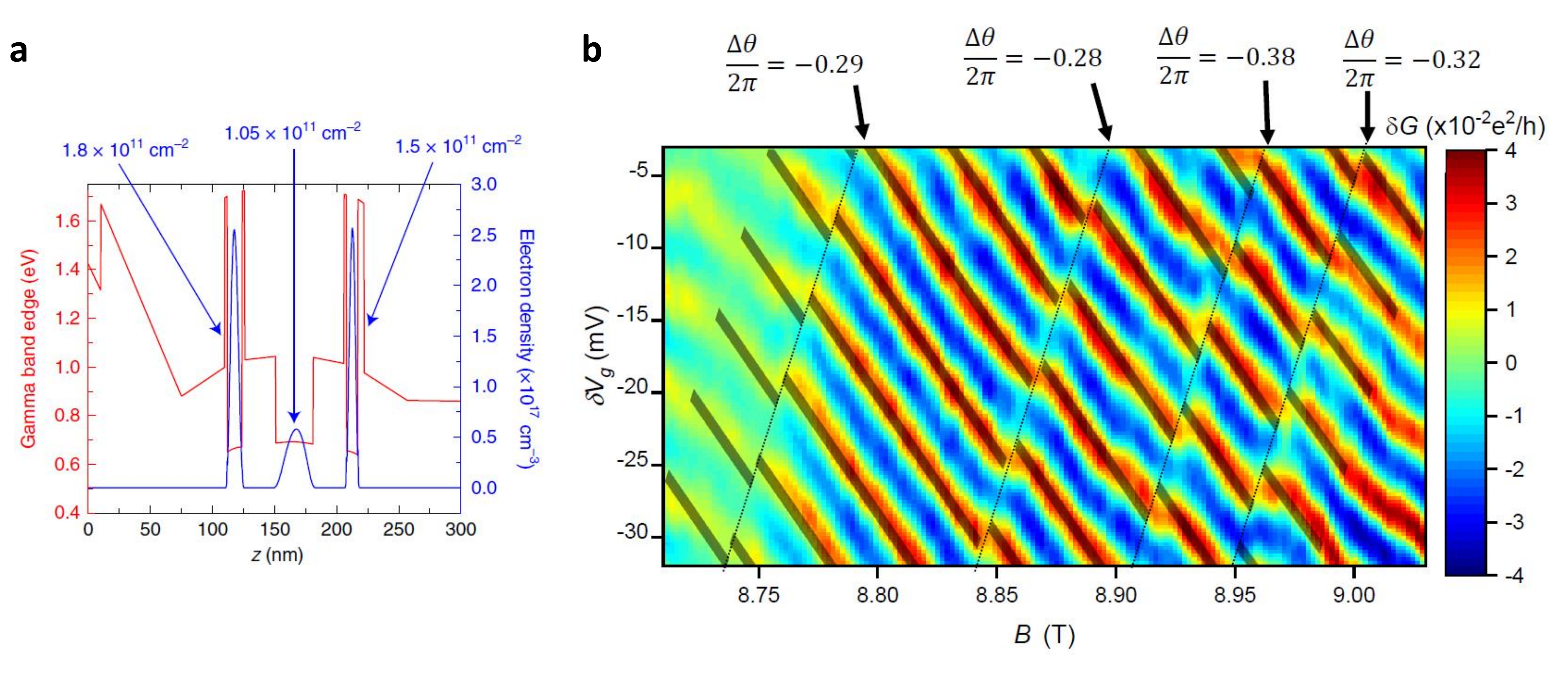}
   \caption{{\bf Detection of anyonic statistics.} \label{fig:fpi3} (a) Conduction band edge (red) and electron density (blue) versus growth direction (z--axis) calculated using a self--consistent Schr\"odinger-Poisson method. The sheet density in each well is indicated.  (b) Conductance oscillations $\delta G$ versus magnetic field $B$ and variation in side gate voltage $\delta V_g$. The predominant behavior shows negatively sloped Aharonov-Bhom interference fringes, with a small number of discrete phase jumps (highlighted by dashed lines). Least-squares fits of $\delta G$ are shown with highlighted stripes, and the extracted changes in phase $\frac{\Delta \theta}{2 \pi}$ are indicated for each discrete jump. { The latter are directly related to the statistical angle of quasiparticles.} Panel a adapted from Ref.~\cite{Nakamura2019}. Panel b adapted from Ref.~\cite{Nakamura2020}.}
\end{figure}  

By applying a finite source-drain voltage $V_{SD}$~\cite{McClure2009, Nakamura2019}, electrons with different energies (and thus different momentum) are injected in the interferometer, inducing an additional phase shift,
\begin{equation}
\delta \theta = \delta \epsilon \frac{\partial k}{\partial \epsilon}L = \frac{\delta \epsilon L}{ v_{edge}} , 
\end{equation}
with $L$ the FPI path length and $v_{edge}$ the edge mode velocity.  With this non-equilibrium configuration, it is possible to infer the speed of edge states by looking at the conductance oscillations as a function of both $V_{SD}$ and gate voltage $V_g$. Such measurements show a characteristic checkerboard lattice with nodes in the interference pattern (Fig.~\ref{fig:fpi2}c) occurring at $V_{SD} = \pm { \pi  v_{edge}}/{eL}$, from which one can extract $v_{edge} = {eL  V_{SD}}/{2\pi }$. Measurements of edge velocity for different Landau levels were reported as a function of filling factor $\nu$, (Fig.~\ref{fig:fpi2}d) ~\cite{Nakamura2019}. The inner, higher index Landau levels were found to generally have lower velocity and correspondingly lower coherence. The general trend observed is an increasing edge velocity {with} decreasing magnetic field (larger filling factor), consistent with the simple picture that edge currents are generated by Hall drift. These observations recall previous work~\cite{Chklovskii1992, Chklovskii1993} that demonstrated that the confining potential is steepest at the outer edge, resulting in a higher electric field and thus higher velocity for the outer edge mode. For higher filling $\nu>2$, these larger velocities are also consistent with the interedge coupling picture, where interacting edge modes are expected to have different velocities~\cite{NgoDinh2012, Ferraro2017, Frigeri2020}, supporting the idea behind the halving of periodicities described above.

These recent results have been obtained in a novel FPI geometry where the sample contains three GaAs wells: a primary quantum well $30$~nm-wide and two additional $13$~nm-thick wells, located on either side of the primary well and separated by thin insulating AlGaAs spacers (Fig.~\ref{fig:fpi3}a). The central well contains the 2D electron gas while the two side wells act as screening layers. {These} fabrication advances allowed the study of FPIs in the extreme AB regime for small interferometer areas of the order of $\sim 1$~$\mu$m$^2$. Structures utilizing screening layers resulted in charging energies, and hence mutual bulk edge coupling, lower by more than one order of magnitude with respect to devices of the same size with conventional gating schemes~\cite{Manfra2014, Sahasrabudhe2018, Nakamura2019, Nakamura2020}. Remarkably, these new devices allowed to observe, for the first time, oscillations with negatively sloped lines of constant phase in the $B-V_g$ plane, hallmark of the AB operating regime, also in the FQH regime. Indeed, by operating the device at high magnetic field, at the center of the $\nu=1/3$ QH plateau, the conductance variation $\delta G$ of the FPI could be studied as a function of $B$ and variation in gate voltage $\delta V_g$. An overall behavior with negatively sloped stripes is clearly visible with a number of discrete phase jumps (Fig.~\ref{fig:fpi3}b) ~\cite{Nakamura2020}. Although not regularly spaced, these jumps have nearly the same slope in the $B-V_g$ plane and have been verified as reproducible in the same study~\cite{Nakamura2020}. Impressively, these findings are fully consistent with Equation (\ref{theta}): the second term on the right hand side predicts a discrete change in phase when the number of localized quasiparticles changes. { These discrete phase jumps are directly reflected in the behaviour of the measured conductance and} constitute the first direct observation of the effect of the anyonic phase contribution $\theta_{\nu}$. Moreover, the observation of a positive slope strongly suggests that these discrete phase jumps are associated with changes in localized quasiparticle number: increasing $B$ is expected to remove quasiparticles from the bulk~\cite{Chamon1997, Stern2008}, while increasing gate voltage would make it electrostatically favorable to increase the number of localized quasiparticles. The authors of Ref.~\cite{Nakamura2020} performed a least-squares fit in the regions between the phase jumps, fitting the conductance data to the form $\delta G = \delta G_0 \cos \left( 2 \pi \frac{1}{3} \frac{B \cdot A}{\phi_0}+\theta_0 \right)$, to determine the value of the change in phase associated with each phase jump. Analyzing  $\theta_0$ across adjacent regions, and under the assumption that each phase jump corresponds to the removal of a localized quasiparticle ($\delta n_q = 1$), they have extracted the value $\theta _{\nu} = 2 \pi \times \left( 0.31 \pm 0.04 \right)$, which well agrees with the predicted value of $1/3$ for the Laughlin state at $\nu=1/3$~\cite{Stern2008, Halperin2011}. To further corroborate their findings, additional measurements varying magnetic field $B$ in a wider range have been performed. {A variation} of the $3\phi_0$ superperiodicity in $\Delta B$ was reported while moving away from the center of the $\nu=1/3$ QH plateau, consistent with the recent prediction of Ref.~\cite{Rosenow2020}. These impressive  measurements provide experimental confirmation for the prediction of fractional braiding statistics at the $\nu = 1/3$ quantum 
Hall state, opening the way for the study of more { complex} FQH states with composite edge physics beyond the Laughlin series.

{ 
\section*{Towards non-Abelian Statistics}
}

The described recent advances inevitably fuel and revitalize the search for signatures of non-Abelian anyons \cite{Nayak2008, Stern2008} in more exotic filling factors such as $\nu=5/2$. Indeed, when the ground state 
of a many-body system is degenerate, the effect of a closed trajectory is not necessarily a mere phase factor. The system can start and end  in different ground states belonging to the same degenerate subspace. Such adiabatic processes may be expressed in terms of a generic unitary transformations within this subspace. Therefore, the topological phase determined by Eq.~(\ref{eq:berry}) is generalized to a matrix rather than a simple phase factor. Braiding quasiparticles within such degenerate subspace allows to build unitary operations which are topologically stable and immune to external perturbations and decoherence \cite{Nayak2008}. Non-Abelian exchange statistics have been predicted for quasiparticles of FQH states such as $\nu=5/2$ or more exotic (and fragile) ones \cite{DasSarma2005, Fendley2005, Freedman2006, Kitaev2006, Nayak2008}. Indeed, the most promising candidates to describe the even-denominator FQH state at $\nu=5/2$ are the Pfaffian (or Moore-Read) \cite{Moore1991, Morf1998} or its conjugate \cite{Chung2003, Levin2007, Stern2008, Bishara2008, Dolev2008, Carrega2011}, which would host non-Abelian anyons. Such quasiparticles have been predicted to possess $e^* =e/4$ fractional charge, and some indications in this direction have been reported by shot noise transport measurements \cite{Chung2003, Dolev2008}. Several models have been put forward \cite{Moore1991,Stern2006,Fendley2007,Levin2007,Bishara2008,Bishara2009}  for the $\nu=5/2$ state, including some without non-Abelian excitations, and from such measurements it is not possible to conclusively determine the non-Abelian nature of this FQH state. { It is worth mentioning that measurements seem to confirm the non-Abelian nature of the $\nu=5/2$ FQH state \cite{Tiemann2012,Banerjee2018}}.

Interferometric setups like Fabry-Perot configurations \cite{Willett2009, 
Willett2019} have been realized at filling $\nu=5/2$. The reported results, mostly in the Coulomb-dominated regime, have been subject of a long debate \cite{Stern2006,Willett2009,Bishara2009,Halperin2011,Willett2013,Willett2019}. Several interpretations of the reported observations have been proposed, but a conclusive answer towards a unique model unequivocally supporting non-Abelian anyons is still lacking. We believe that recent technological advances, such as the screening layer mechanism \cite{Nakamura2019,Nakamura2020} and the anyon collider platform \cite{Bocquillon2014,RosenowPRL2016,Roussel2017,Glattli2017,Bauerle2018,Bartolomei2020}, would be  beneficial for the inspection of $\nu=5/2$ or other exotic filling factors, allowing to clearly enter the AB regime, where evidence of statistical angle should be unambiguous and could constitute a smoking gun for such longstanding detection of non-Abelian quasiparticles and their ensuing manipulation.

\section*{Discussion}

{  The two interferometric setups described in the previous sections clearly share differences and similarities, complementarities and incompatibilities, pros and cons. {  Due to its open geometry, one of the advantages of the MZI scheme is the absence of Coulomb blockade physics that strongly influences the FPI.} The success of the FPI in demonstrating AB oscillations and clear signatures of the fractional statistics in no way takes away importance from the studies performed on MZI. Moreover, preliminary results on setups that avoid the subtleties of the Corbino geometry \cite{Deviatov2012} suggest possible alternative routes for the design and construction of a MZI able to show signatures of fractional statistics. Further investigations may reveal the key to a {  MZI working in the FQH regime} and allow full exploitation of a setup that is free from Coulomb blockade effects, whose control have been proven crucial for the success of the FPI.}

{  Very recently, novel heterostructures based on high-quality graphene devices have been demonstrated to be an interesting and alternative platform for electronic interferometry \cite{Wei2017,Deprez2021,Ronen2021,Jo2021}. By properly patterning several split-gates, it has been possible to realize FPIs on graphene devices free of charging effects, observing clear evidence of AB oscillations in the integer QH regime. These promising results, and the reported long coherence length, will allow for a new generation of electronic interferometers with 2D materials, in which several FPIs can be easily joined, in order to realize possible braiding schemes both at integer and fractional filling factors.}

{  Finally, remarkable results have been obtained exploiting time-dependent signals. An entirely complementary field of electron quantum optics \cite{Bocquillon2014,Ferraro2017a,Roussel2017,Glattli2017,Bauerle2018} has been developed in parallel to the standard interferometry. These platforms allow to exploit two-particle interference like the Hanbury-Brown-Twiss \cite{Oliver1999,Henny1999,Oberholzer2000,Safi2001,Samuelsson2004,Campagnano2012,Campagnano2013} and Hong-Ou-Mandel \cite{Jonckheere2012,Freulon2015} schemes, that are directly sensitive to two-particle statistics and paved the way to the recent success of the anyon collider experiments \cite{RosenowPRL2016,Bartolomei2020}. Furthermore, anyon colliders and interferometric setups could be merged and joined together, offering the opportunity to exploit complementarities and reciprocal advantages, and develop novel platforms for topological quantum computing.}

\section*{Acknowledgements}
This activity was partially supported by the SUPERTOP project, QUANTERA ERA-NET Cofound in Quantum Technologies, and by the FET-OPEN project AndQC. L.~C.~acknowledges funding by the EU Marie Curie Global fellowship TOPOCIRCUS-841894 - Simulations of Topological Phases in Superconducting Circuits.

\section{BOX: Optical interferometers}
\label{Box1}

The Mach-Zehnder interferometer (MZI) is a two-path interferometric setup originally aimed at determining a relative phase shift between two collimated beams originating from a single source (panel a). The device is named after Ludwig Mach and Ludwig Zehnder, who proposed the interferometer in 1891, { and it has many applications, from aerodynamics and plasma physics to gravitational wave detection}. An incoming beam of photons from source S is split at the first beam-splitter BS1, and the two outgoing beams follow different paths via mirrors M1 and M2. The beams rejoin at BS2, and interference as a function of their relative phase $\theta$ acquired along the two different paths is observed at the detectors D1 and D2. Photons are generated with a precise momentum $k$, and their relative phase is due to the dynamical contribution $\theta_L = k \Delta L$ that is accumulated due to propagation along paths of different length $\Delta L$. 

The Fabry-Perot interferometer (FPI) is a resonator originally constituted by two partially reflecting optical flats placed one in front of the other, with the reflecting surfaces facing each other (panel b). It is named after Charles Fabry and Alfred Perot, who developed the instrument in 1899. The optical flats act as beam-splitters. Beam-splitters are rotated by $\pi/2$ with respect to the MZI. A photon incoming from source S impinges on the first beam-splitter BS1, where it can be either reflected (and reach detector D1) or transmitted. In the second case, it propagates for a given distance $L$ that separates the first beam-splitter from the second, BS2, where it can be either transmitted and reach detector D2, or reflected back to BS1. This sequence can repeat an arbitrary number of times, and the outgoing signals at D1 or D2 are the result of the coherent sum of all amplitudes for multiple bouncing between the two beam-splitters. A sharp resonance in the transmitted signal at D2 occurs when the light exhibits constructive interference after one round trip, at photon wavelength matching twice the distance between the mirrors, $\lambda=2L$.  FPIs are widely used as resonators, spectrometers, optical filters, and interferometers in optics, telecommunications, astronomy, and gravitational wave detection.

\begin{figure}[h!]
   \includegraphics[width=0.8\textwidth]{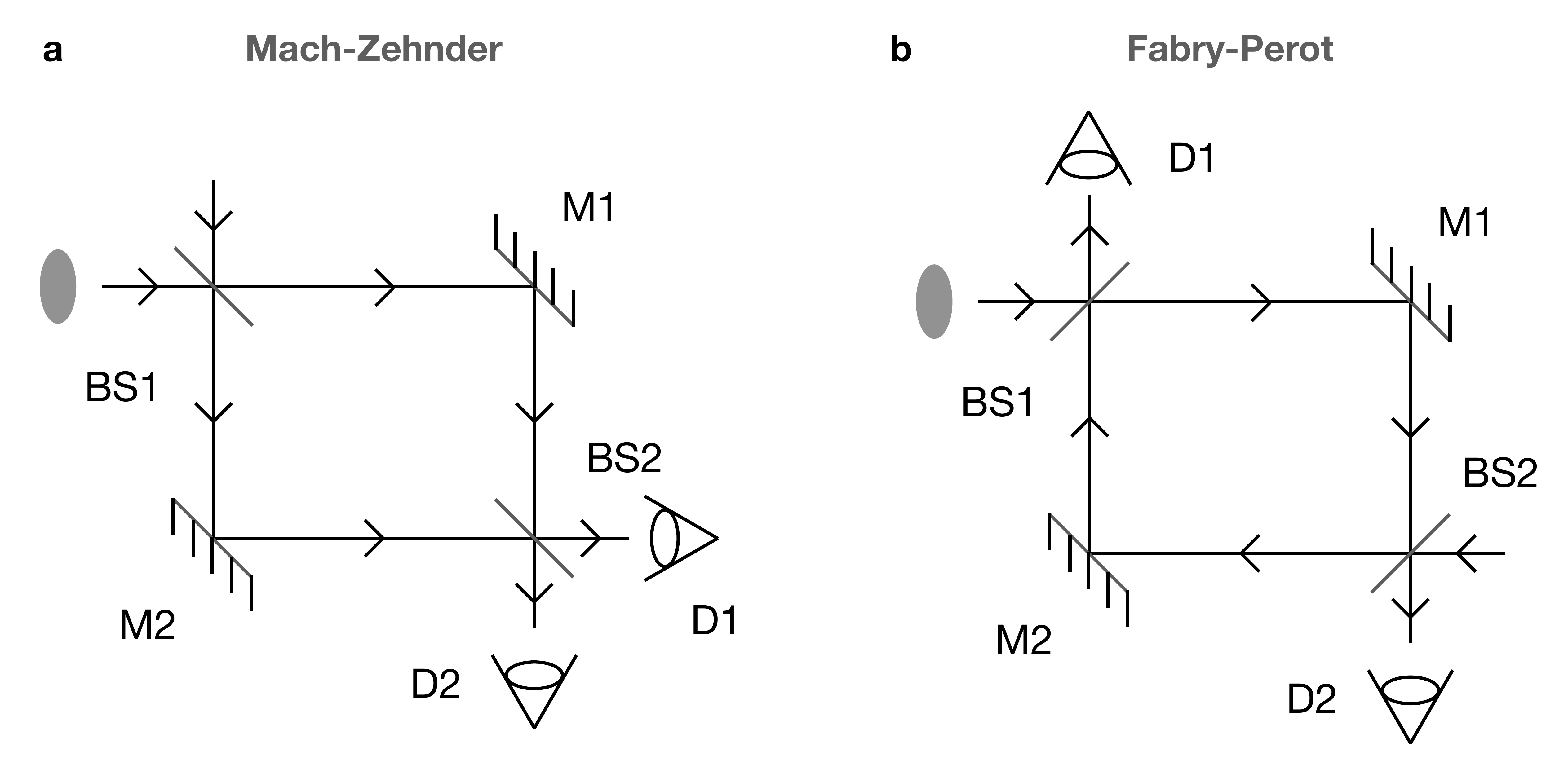}
\end{figure}

\section{BOX: Quantum Point Contacts}
\label{Box2}

Starting from a 2DEG, a further restriction of carrier motion to only one dimension can be obtained by using a split-gate geometry, where a small channel region is defined by two gates, separated by a few hundreds of nanometers. By applying a negative gate voltage, these gates tailor the width of the channel by electrostatically depleting the 2DEG under and next to them. This leaves only a small opening in the 2DEG for the electrons to pass through, as sketched in the figure, where the part of the 2DEG depleted by the gates is indicated in gray. If a large negative gate voltage is applied, the gates can even completely close the channel (pinch-off). Such device is called a quantum point contact (QPC).

When the Fermi wavelength $\lambda_F$ is comparable to the width of the QPC, the transverse momentum becomes quantized, forming a series of parabolic one-dimensional subbands \cite{Beenakker,Landauer,Landauer2, Conductance1,Conductance2,Schapers}. In a 2DEG in the QH regime, current is carried by chiral edge channels, which at the two opposite sides of the Hall-bar propagate in opposite directions. If a QPC is inserted in the device, by squeezing such constriction, the counter-propagating edge channels are forced to get closer. If their distance becomes comparable to the magnetic length, they will start to interact and (partially) be backscattered. The first backscattered edge states are the ones formed by the highest Landau levels, because they travel farther from the edges. After, edge states with lower Landau index are backscattered, until all edge channels are backscattered (pinch-off). The figure illustrates this for $\nu = 2$ in the bulk. Far from the constriction, the two edge channels propagate on either side of the Hall-bar, but the inner one (red) is backscattered by the QPC, so only one (the outer) channel passes through it (blue). Therefore the filling factor in the constriction is $\nu_{\text{QPC}}=1$.

\begin{figure}[h!]
   \includegraphics[width=0.8\textwidth]{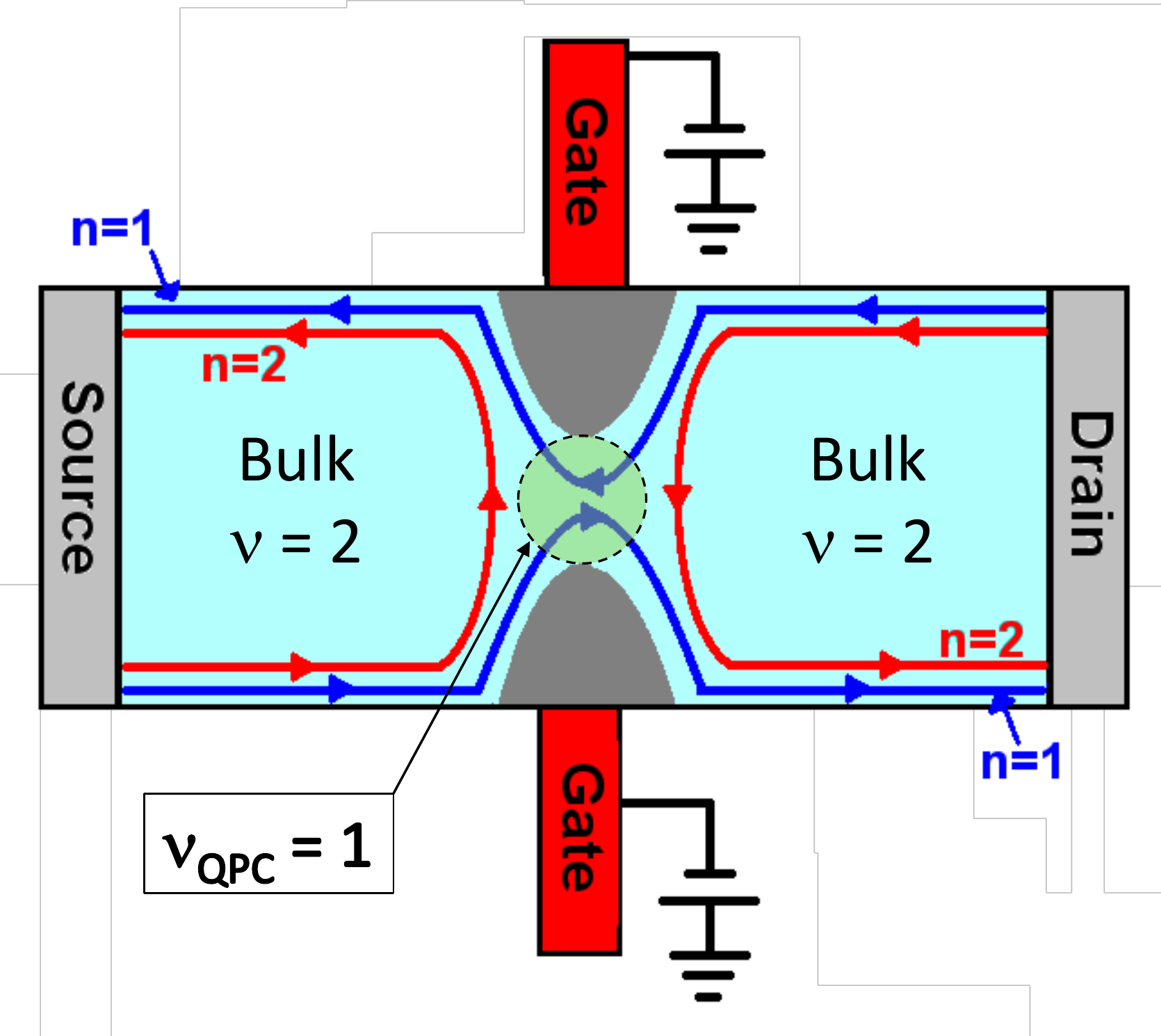}
\end{figure} 

\section{BOX: Counter-propagating edge states and neutral modes}
\label{Box3}

The complexity of edge reconstruction in the FQH effect can be visualized by considering the prototypical hole-like $\nu=2/3$ state. The latter has been described as the formation of a $\nu=1/3$ Hall droplet by depletion  of a $\nu=1$ state in the bulk \cite{MacDonald1990,Wen1990,Wen1990a}. Edge excitations are described by the coexistence of a $\nu=1$ forward propagating (downstream) and a $\nu=1/3$ backward (counter-propagating or upstream) edge mode (panel a). However, this picture results in a value of the Hall conductance $G_H = \left( 1 + \frac{1}{3} \right) e^2/h$ that is different from the experimentally observed universal value of $\frac{2}{3} \, e^2/h$. Furthermore, the presence of an upstream charge flow is at odds with the absence of any experimental signature in this sense. The key observation is that counter-propagating charge modes, besides Coulomb interactions, are sensitive to the presence of disorder (panel b). Interchannel interactions can be considered in a dipole scheme by introducing "charge" ($c$) and "neutral" ($n$) modes (panel c). However, the resulting conductance still has a nonuniversal value
\begin{equation*}
G_H = \Delta \frac{2}{3} \frac{e^2}{h},
\end{equation*}
where $\Delta\geq 1$ depends on the velocities of the modes and the Coulomb interaction strength.

 It has been argued that a correct inclusion of disorder-induced scattering between the two edge modes is necessary to reconcile with the proper quantized value of $G_H$ \cite{Kane1994,Kane1995}. The crucial point is that the counter-propagating nature of the two charge modes allows for backscattering mediated by disorder, which is a relevant interaction between the two channels. Physically, subtraction of one electron from the forward channel is accompanied by the addition of 3 quasiparticles to the backward channel, so that tunneling between the two modes effectively involves only the neutral mode, as emphasized in panel~c. The residual interaction between charge and neutral mode can be shown to be irrelevant in the renormalization group sense, leading to {  the fixed point} $\Delta\to 1$, eventually corresponding to the experimentally observed conductance $G_H = \frac{2}{3} \, e^2/h$ \cite{Kane1994,Braggio2012}.

Notably, the $\nu=2/3$ edge state has recently been experimentally synthesized  by realizing a junction between $\nu=1$ and $\nu=1/3$ Hall droplets, bringing  the counter-propagating associated edge states into close proximity. The experiment demonstrated that equilibration between the two states results in a  $G_H = \frac{2}{3} \, e^2/h$ Hall conductance, whereas in absence of scattering a value of  $G_H = \frac{4}{3} \, e^2/h$ is observed \cite{Cohen2019}. {  At the same time, we note that the conditions needed to approach the low energy fixed point may be demanding, and a robust universal value of 2/3 is alternatively predicted in presence of edge equilibration involving also decoherence \cite{Nosiglia2018,Protopopov2017}}.

\begin{figure}[h!]
   \includegraphics[width=0.8\textwidth]{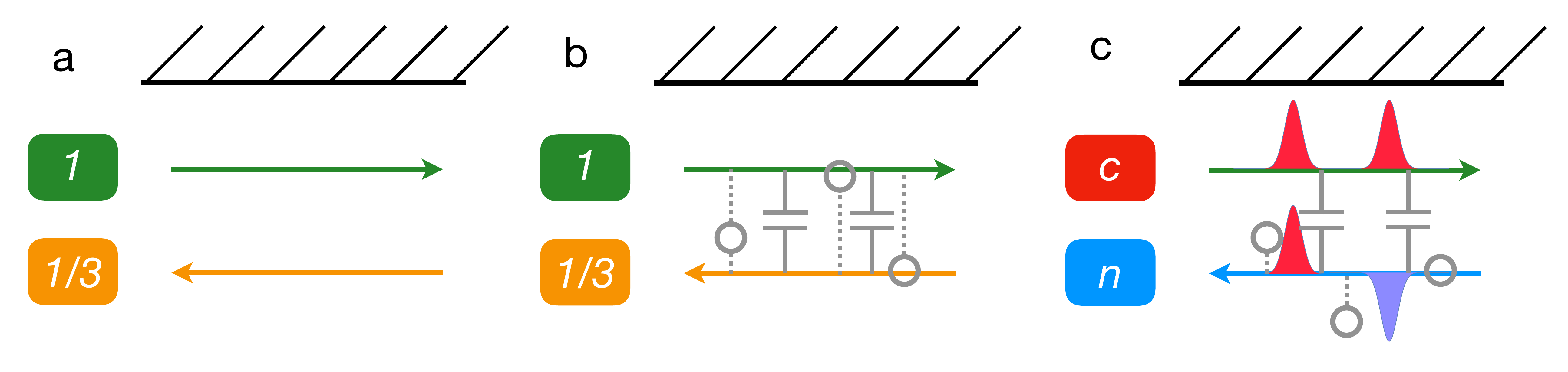}
\end{figure} 

\bibliographystyle{ieeetr}
\bibliography{biblio}

\begin{thebibliography}{100}

\bibitem{vonkl2020}
K.~von Klitzing, T.~Chakraborty, P.~Kim, V.~Madhavan, X.~Dai, J.~McIver,
  Y.~Tokura, L.~Savary, D.~Smirnova, A.~M. Rey, C.~Felser, J.~Gooth, and X.~Qi,
  ``{40 years of the quantum Hall effect},'' {\em Nature Reviews Physics},
  vol.~2, no.~8, pp.~397--401, 2020.

\bibitem{TKNN}
D.~J. Thouless, M.~Kohmoto, M.~P. Nightingale, and M.~den Nijs, ``{Quantized
  Hall Conductance in a Two-Dimensional Periodic Potential},'' {\em Phys. Rev.
  Lett.}, vol.~49, pp.~405--408, Aug 1982.

\bibitem{Haldane-NobelLecture}
F.~D.~M. Haldane, ``Nobel lecture: Topological quantum matter,'' {\em Rev. Mod.
  Phys.}, vol.~89, p.~040502, Oct 2017.

\bibitem{Stern2008}
A.~Stern, ``{Anyons and the quantum Hall effect---A pedagogical review},'' {\em
  Ann. Phys.}, vol.~323, no.~1, pp.~204--249, 2008.

\bibitem{Nayak2008}
C.~Nayak, S.~H. Simon, A.~Stern, M.~Freedman, and S.~Das~Sarma, ``{Non-Abelian
  anyons and topological quantum computation},'' {\em {Rev. Mod. Phys.}},
  vol.~80, pp.~1083--1159, Sep 2008.

\bibitem{Tsui1982}
D.~C. Tsui, H.~L. Stormer, and A.~C. Gossard, ``{Two-Dimensional
  Magnetotransport in the Extreme Quantum Limit},'' {\em Phys. Rev. Lett.},
  vol.~48, pp.~1559--1562, May 1982.

\bibitem{Laughlin1983}
R.~B. Laughlin, ``{Anomalous Quantum Hall Effect: An Incompressible Quantum
  Fluid with Fractionally Charged Excitations},'' {\em Phys. Rev. Lett.},
  vol.~50, pp.~1395--1398, May 1983.

\bibitem{Bocquillon2014}
E.~Bocquillon, V.~Freulon, F.~D. Parmentier, J.-M. Berroir, B.~Pla{\c c}ais,
  C.~Wahl, J.~Rech, T.~Jonckheere, T.~Martin, C.~Grenier, D.~Ferraro,
  P.~Degiovanni, and G.~F{\`e}ve, ``Electron quantum optics in ballistic chiral
  conductors,'' {\em Ann. Phys.}, vol.~526, no.~1-2, pp.~1--30, 2014.

\bibitem{RosenowPRL2016}
B.~Rosenow, I.~P. Levkivskyi, and B.~I. Halperin, ``Current correlations from a
  mesoscopic anyon collider,'' {\em Phys. Rev. Lett.}, vol.~116, p.~156802, Apr
  2016.

\bibitem{Roussel2017}
B.~Roussel, C.~Cabart, G.~F{\`e}ve, E.~Thibierge, and P.~Degiovanni, ``Electron
  quantum optics as quantum signal processing,'' {\em Phys. Status Solidi B},
  vol.~254, no.~3, p.~1600621, 2017.

\bibitem{Glattli2017}
D.~C. Glattli and P.~S. Roulleau, ``Levitons for electron quantum optics,''
  {\em Phys. Status Solidi B}, vol.~254, no.~3, p.~1600650, 2017.

\bibitem{Bauerle2018}
C.~B{\"a}uerle, D.~C. Glattli, T.~Meunier, F.~Portier, P.~Roche, P.~Roulleau,
  S.~Takada, and X.~Waintal, ``Coherent control of single electrons: a review
  of current progress,'' {\em Reports on Progress in Physics}, vol.~81,
  p.~056503, apr 2018.

\bibitem{Ionicioiu2001}
R.~Ionicioiu, G.~Amaratunga, and F.~Udrea, ``Quantum computation with ballistic
  electrons,'' {\em Internat. J. Modern Phys. B}, vol.~15, no.~02,
  pp.~125--133, 2001.

\bibitem{Stace2004}
T.~M. Stace, C.~H.~W. Barnes, and G.~J. Milburn, ``{Mesoscopic One-Way Channels
  for Quantum State Transfer via the Quantum Hall Effect},'' {\em Phys. Rev.
  Lett.}, vol.~93, p.~126804, Sep 2004.

\bibitem{Feve2008}
G.~F\`eve, P.~Degiovanni, and T.~Jolicoeur, ``{Quantum detection of electronic
  flying qubits in the integer quantum Hall regime},'' {\em Phys. Rev. B},
  vol.~77, p.~035308, Jan 2008.

\bibitem{Giovannetti2008}
V.~Giovannetti, F.~Taddei, D.~Frustaglia, and R.~Fazio, ``{Multichannel
  architecture for electronic quantum Hall interferometry},'' {\em {Phys. Rev.
  B}}, vol.~77, p.~155320, Apr 2008.

\bibitem{Bordone2019}
P.~Bordone, L.~Bellentani, and A.~Bertoni, ``{Quantum computing with
  quantum-Hall edge state interferometry},'' {\em Semicond Sci. Technol.},
  vol.~34, p.~103001, sep 2019.

\bibitem{Shimizu2020}
T.~Shimizu, T.~Nakamura, Y.~Hashimoto, A.~Endo, and S.~Katsumoto,
  ``{Gate-controlled unitary operation on flying spin qubits in quantum Hall
  edge states},'' {\em Phys. Rev. B}, vol.~102, p.~235302, Dec 2020.

\bibitem{Leinaas1977}
J.~M. Leinaas and J.~Myrheim, ``On the theory of identical particles,'' {\em Il
  Nuovo Cimento B}, vol.~37, no.~1, pp.~1--23, 1977.

\bibitem{Wilczek1982}
F.~Wilczek, ``{Quantum Mechanics of Fractional-Spin Particles},'' {\em Phys.
  Rev. Lett.}, vol.~49, pp.~957--959, Oct 1982.

\bibitem{Halperin2020}
B.~I. Halperin and J.~K. Jain, {\em Fractional Quantum Hall Effects}.
\newblock WORLD SCIENTIFIC, 2020.

\bibitem{Nakamura2020}
J.~Nakamura, S.~Liang, G.~C. Gardner, and M.~J. Manfra, ``Direct observation of
  anyonic braiding statistics,'' {\em Nat. Phys.}, vol.~16, pp.~931--936, 2020.

\bibitem{Bartolomei2020}
H.~Bartolomei, M.~Kumar, R.~Bisognin, A.~Marguerite, J.-M. Berroir,
  E.~Bocquillon, B.~Pla{\c c}ais, A.~Cavanna, Q.~Dong, U.~Gennser, Y.~Jin, and
  G.~F{\`e}ve, ``Fractional statistics in anyon collisions,'' {\em Science},
  vol.~368, no.~6487, pp.~173--177, 2020.

\bibitem{Wilczek1990}
F.~Wilczek, {\em {Fractional Statistics and Anyon Superconductivity}}.
\newblock World Scientific, 1990.

\bibitem{Berry1984}
M.~V. Berry, ``Quantal phase factors accompanying adiabatic changes,'' {\em
  Proceedings of the Royal Society of London. A. Mathematical and Physical
  Sciences}, vol.~392, no.~1802, pp.~45--57, 1984.

\bibitem{Aharonov1959}
Y.~Aharonov and D.~Bohm, ``{Significance of Electromagnetic Potentials in the
  Quantum Theory},'' {\em Phys. Rev.}, vol.~115, pp.~485--491, Aug 1959.

\bibitem{Wen1995}
X.-G. Wen, ``{Topological orders and edge excitations in fractional quantum
  Hall states},'' {\em Adv. Phys.}, vol.~44, no.~5, pp.~405--473, 1995.

\bibitem{Arovas1984}
D.~Arovas, J.~R. Schrieffer, and F.~Wilczek, ``{Fractional Statistics and the
  Quantum Hall Effect},'' {\em Phys. Rev. Lett.}, vol.~53, pp.~722--723, Aug
  1984.

\bibitem{dePicciotto1997}
R.~de~Picciotto, M.~Reznikov, M.~Heiblum, V.~Umansky, G.~Bunin, and D.~Mahalu,
  ``Direct observation of a fractional charge,'' {\em Nature}, vol.~389,
  p.~162, 1997.

\bibitem{Saminadayar1997}
L.~Saminadayar, D.~C. Glattli, Y.~Jin, and B.~Etienne, ``{Observation of the
  $\mathit{e}\mathit{/}3$ Fractionally Charged Laughlin Quasiparticle},'' {\em
  Phys. Rev. Lett.}, vol.~79, pp.~2526--2529, Sep 1997.

\bibitem{Ji2003}
Y.~Ji, Y.~Chung, D.~Sprinzak, M.~Heiblum, D.~Mahalu, and H.~Shtrikman, ``{An
  electronic Mach-Zehnder interferometer},'' {\em {Nature}}, vol.~422, p.~415,
  2003.

\bibitem{Marquardt2004}
F.~Marquardt and C.~Bruder, ``{Influence of Dephasing on Shot Noise in an
  Electronic Mach-Zehnder Interferometer},'' {\em Phys. Rev. Lett.}, vol.~92,
  p.~056805, Feb 2004.

\bibitem{Chung2005}
V.~S.-W. Chung, P.~Samuelsson, and M.~B\"uttiker, ``{Visibility of current and
  shot noise in electrical Mach-Zehnder and Hanbury Brown Twiss
  interferometers},'' {\em Phys. Rev. B}, vol.~72, p.~125320, Sep 2005.

\bibitem{Forster2005}
H.~F\"orster, S.~Pilgram, and M.~B\"uttiker, ``{Decoherence and full counting
  statistics in a Mach-Zehnder interferometer},'' {\em Phys. Rev. B}, vol.~72,
  p.~075301, Aug 2005.

\bibitem{Neder2006a}
I.~Neder, M.~Heiblum, Y.~Levinson, D.~Mahalu, and V.~Umansky, ``{Unexpected
  Behavior in a Two-Path Electron Interferometer},'' {\em {Phys. Rev. Lett.}},
  vol.~96, p.~016804, Jan 2006.

\bibitem{Neder2007c}
I.~Neder, F.~Marquardt, M.~Heiblum, D.~Mahalu, and V.~Umansky, ``Controlled
  dephasing of electrons by non-gaussian shot noise,'' {\em Nat. Phys.},
  vol.~3, pp.~534--537, jun 2007.

\bibitem{Neder2007d}
I.~Neder, M.~Heiblum, D.~Mahalu, and V.~Umansky, ``{Entanglement, Dephasing,
  and Phase Recovery via Cross-Correlation Measurements of Electrons},'' {\em
  {Phys. Rev. Lett.}}, vol.~98, p.~036803, Jan 2007.

\bibitem{Litvin2007}
L.~V. Litvin, H.~P. Tranitz, W.~Wegscheider, and C.~Strunk, ``{Decoherence and
  single electron charging in an electronic Mach-Zehnder interferometer},''
  {\em {Phys. Rev. B}}, vol.~75, p.~033315, 2007.

\bibitem{Roulleau2007}
P.~Roulleau, F.~Portier, D.~C. Glattli, P.~Roche, A.~Cavanna, G.~Faini,
  U.~Gennser, and D.~Mailly, ``{Finite bias visibility of the electronic
  Mach-Zehnder interferometer},'' {\em {Phys. Rev. B}}, vol.~76, p.~161309(R),
  2007.

\bibitem{Litvin2008a}
L.~V. Litvin, A.~Helzel, H.-P. Tranitz, W.~Wegscheider, and C.~Strunk, ``{Two
  beam Aharonov--Bohm interference in the integer quantum Hall regime},'' {\em
  Physica E}, vol.~40, no.~5, p.~1706, 2008.

\bibitem{Litvin2008}
L.~V. Litvin, A.~Helzel, H.-P. Tranitz, W.~Wegscheider, and C.~Strunk,
  ``{Edge-channel interference controlled by Landau level filling},'' {\em
  {Phys. Rev. B}}, vol.~78, p.~075303, Aug 2008.

\bibitem{Roulleau2008}
P.~Roulleau, F.~Portier, P.~Roche, A.~Cavanna, G.~Faini, U.~Gennser, and
  D.~Mailly, ``{Direct Measurement of the Coherence Length of Edge States in
  the Integer Quantum Hall Regime},'' {\em {Phys. Rev. Lett.}}, vol.~100,
  p.~126802, Mar 2008.

\bibitem{Roulleau2008a}
P.~Roulleau, F.~Portier, P.~Roche, A.~Cavanna, G.~Faini, U.~Gennser, and
  D.~Mailly, ``{Noise Dephasing in Edge States of the Integer Quantum Hall
  Regime},'' {\em {Phys. Rev. Lett.}}, vol.~101, p.~186803, Oct 2008.

\bibitem{Roulleau2008b}
P.~Roulleau, F.~Portier, D.~C. Glattli, A.~Cavanna, G.~Faini, U.~Gennser,
  D.~Mailly, and P.~Roche, ``{High visibility in an electronic Mach--Zehnder
  interferometer with random phase fluctuations},'' {\em Physica E}, vol.~40,
  no.~5, p.~1048, 2008.

\bibitem{Roulleau2009}
P.~Roulleau, F.~Portier, P.~Roche, A.~Cavanna, G.~Faini, U.~Gennser, and
  D.~Mailly, ``{Tuning Decoherence with a Voltage Probe},'' {\em {Phys. Rev.
  Lett.}}, vol.~102, p.~236802, Jun 2009.

\bibitem{Bieri2009}
E.~Bieri, M.~Weiss, O.~G\"oktas, M.~Hauser, C.~Sch\"onenberger, and
  S.~Oberholzer, ``{Finite-bias visibility dependence in an electronic
  Mach-Zehnder interferometer},'' {\em {Phys. Rev. B}}, vol.~79, p.~245324, Jun
  2009.

\bibitem{Litvin2010}
L.~V. Litvin, A.~Helzel, H.-P. Tranitz, W.~Wegscheider, and C.~Strunk, ``{Phase
  of the transmission amplitude for a quantum dot embedded in the arm of an
  electronic Mach-Zehnder interferometer},'' {\em {Phys. Rev. B}}, vol.~81,
  p.~205425, May 2010.

\bibitem{Weisz2012}
E.~Weisz, H.~K. Choi, M.~Heiblum, Y.~Gefen, V.~Umansky, and D.~Mahalu,
  ``{Controlled Dephasing of an Electron Interferometer with a Path Detector at
  Equilibrium},'' {\em {Phys. Rev. Lett.}}, vol.~109, p.~250401, Dec 2012.

\bibitem{Levkivskyi2008}
I.~P. Levkivskyi and E.~V. Sukhorukov, ``{Dephasing in the electronic
  Mach-Zehnder interferometer at filling factor $\ensuremath{\nu}=2$},'' {\em
  {Phys. Rev. B}}, vol.~78, p.~045322, Jul 2008.

\bibitem{Levkivskyi2009}
I.~P. Levkivskyi and E.~V. Sukhorukov, ``{Noise-Induced Phase Transition in the
  Electronic Mach-Zehnder Interferometer},'' {\em Phys. Rev. Lett.}, vol.~103,
  p.~036801, Jul 2009.

\bibitem{Rosenow2012}
B.~Rosenow and Y.~Gefen, ``{Dephasing by a Zero-Temperature Detector and the
  Friedel Sum Rule},'' {\em {Phys. Rev. Lett.}}, vol.~108, p.~256805, Jun 2012.

\bibitem{Helzel2015}
A.~Helzel, L.~V. Litvin, I.~P. Levkivskyi, E.~V. Sukhorukov, W.~Wegscheider,
  and C.~Strunk, ``{Counting statistics and dephasing transition in an
  electronic Mach-Zehnder interferometer},'' {\em Phys. Rev. B}, vol.~91,
  p.~245419, Jun 2015.

\bibitem{Chalker2007}
J.~T. Chalker, Y.~Gefen, and M.~Y. Veillette, ``{Decoherence and interactions
  in an electronic Mach-Zehnder interferometer},'' {\em Phys. Rev. B}, vol.~76,
  p.~085320, Aug 2007.

\bibitem{Huynh2012}
P.-A. Huynh, F.~Portier, H.~le~Sueur, G.~Faini, U.~Gennser, D.~Mailly,
  F.~Pierre, W.~Wegscheider, and P.~Roche, ``{Quantum Coherence Engineering in
  the Integer Quantum Hall Regime},'' {\em {Phys. Rev. Lett.}}, vol.~108,
  p.~256802, Jun 2012.

\bibitem{Duprez2019}
H.~Duprez, E.~Sivre, A.~Anthore, A.~Aassime, A.~Cavanna, A.~Ouerghi,
  U.~Gennser, and F.~Pierre, ``{Macroscopic Electron Quantum Coherence in a
  Solid-State Circuit},'' {\em {Phys. Rev. X}}, vol.~9, p.~021030, May 2019.

\bibitem{Chirolli2012}
L.~Chirolli, D.~Venturelli, F.~Taddei, R.~Fazio, and V.~Giovannetti,
  ``{Proposal for a Datta-Das transistor in the quantum Hall regime},'' {\em
  {Phys. Rev. B}}, vol.~85, p.~155317, Apr 2012.

\bibitem{Chirolli2013}
L.~Chirolli, F.~Taddei, R.~Fazio, and V.~Giovannetti, ``{Interactions in
  Electronic Mach-Zehnder Interferometers with Copropagating Edge Channels},''
  {\em {Phys. Rev. Lett.}}, vol.~111, p.~036801, Jul 2013.

\bibitem{Karmakar2015}
B.~Karmakar, D.~Venturelli, L.~Chirolli, V.~Giovannetti, R.~Fazio, S.~Roddaro,
  L.~N. Pfeiffer, K.~W. West, F.~Taddei, and V.~Pellegrini, ``{Nanoscale
  Mach-Zehnder interferometer with spin-resolved quantum Hall edge states},''
  {\em {Phys. Rev. B}}, vol.~92, p.~195303, Nov 2015.

\bibitem{Deviatov2011}
E.~V. Deviatov, A.~Ganczarczyk, A.~Lorke, G.~Biasiol, and L.~Sorba, ``{Quantum
  Hall Mach-Zehnder interferometer far beyond equilibrium},'' {\em {Phys. Rev.
  B}}, vol.~84, p.~235313, Dec 2011.

\bibitem{Deviatov2012}
E.~V. Deviatov, S.~V. Egorov, G.~Biasiol, and L.~Sorba, ``{Quantum Hall
  Mach-Zehnder interferometer at fractional filling factors},'' {\em {EPL}
  (Europhysics Letters)}, vol.~100, p.~67009, dec 2012.

\bibitem{Chklovskii1992}
D.~B. Chklovskii, B.~I. Shklovskii, and L.~I. Glazman, ``Electrostatics of edge
  channels,'' {\em {Phys. Rev. B}}, vol.~46, pp.~4026--4034, Aug 1992.

\bibitem{Chklovskii1993}
D.~B. Chklovskii, K.~A. Matveev, and B.~I. Shklovskii, ``{Ballistic conductance
  of interacting electrons in the quantum Hall regime},'' {\em {Phys. Rev. B}},
  vol.~47, pp.~12605--12617, May 1993.

\bibitem{Paradiso2011}
N.~Paradiso, S.~Heun, S.~Roddaro, D.~Venturelli, F.~Taddei, V.~Giovannetti,
  R.~Fazio, G.~Biasiol, L.~Sorba, and F.~Beltram, ``{Spatially resolved
  analysis of edge-channel equilibration in quantum Hall circuits},'' {\em
  Phys. Rev. B}, vol.~83, p.~155305, Apr 2011.

\bibitem{Paradiso2012}
N.~Paradiso, S.~Heun, S.~Roddaro, L.~Sorba, F.~Beltram, G.~Biasiol, L.~N.
  Pfeiffer, and K.~W. West, ``{Imaging Fractional Incompressible Stripes in
  Integer Quantum Hall Systems},'' {\em Phys. Rev. Lett.}, vol.~108, p.~246801,
  Jun 2012.

\bibitem{Wei2017}
D.~S. Wei, T.~van~der Sar, J.~D. Sanchez-Yamagishi, K.~Watanabe, T.~Taniguchi,
  P.~Jarillo-Herrero, B.~I. Halperin, and A.~Yacoby, ``{Mach-Zehnder
  interferometry using spin- and valley-polarized quantum Hall edge states in
  graphene},'' {\em Science Advances}, vol.~3, no.~8, p.~e1700600, 2017.

\bibitem{Jo2021}
M.~Jo, P.~Brasseur, A.~Assouline, G.~Fleury, H.-S. Sim, K.~Watanabe,
  T.~Taniguchi, W.~Dumnernpanich, P.~Roche, D.~C. Glattli, N.~Kumada, F.~D.
  Parmentier, and P.~Roulleau, ``Quantum hall valley splitters and a tunable
  mach-zehnder interferometer in graphene,'' {\em Phys. Rev. Lett.}, vol.~126,
  p.~146803, Apr 2021.

\bibitem{Amet2014}
F.~Amet, J.~R. Williams, K.~Watanabe, T.~Taniguchi, and D.~Goldhaber-Gordon,
  ``{Selective Equilibration of Spin-Polarized Quantum Hall Edge States in
  Graphene},'' {\em Phys. Rev. Lett.}, vol.~112, p.~196601, May 2014.

\bibitem{Zimmermann2017}
K.~Zimmermann, A.~Jordan, F.~Gay, K.~Watanabe, T.~Taniguchi, Z.~Han,
  V.~Bouchiat, H.~Sellier, and B.~Sac{\'e}p{\'e}, ``{Tunable transmission of
  quantum Hall edge channels with full degeneracy lifting in split-gated
  graphene devices},'' {\em Nature Communications}, vol.~8, no.~1, p.~14983,
  2017.

\bibitem{Idrisov2018}
E.~G. Idrisov, I.~P. Levkivskyi, and E.~V. Sukhorukov, ``{Dephasing in a
  Mach-Zehnder Interferometer by an Ohmic Contact},'' {\em Phys. Rev. Lett.},
  vol.~121, p.~026802, Jul 2018.

\bibitem{Duprez2019transmitting}
H.~Duprez, E.~Sivre, A.~Anthore, A.~Aassime, A.~Cavanna, U.~Gennser, and
  F.~Pierre, ``{Transmitting the quantum state of electrons across a metallic
  island with Coulomb interaction},'' {\em Science}, vol.~366, no.~6470,
  pp.~1243--1247, 2019.

\bibitem{Gurman2016}
I.~Gurman, R.~Sabo, M.~Heiblum, V.~Umansky, and D.~Mahalu, ``Dephasing of an
  electronic two-path interferometer,'' {\em {Phys. Rev. B}}, vol.~93,
  p.~121412, Mar 2016.

\bibitem{Bhattacharyya2019}
R.~Bhattacharyya, M.~Banerjee, M.~Heiblum, D.~Mahalu, and V.~Umansky,
  ``{Melting of Interference in the Fractional Quantum Hall Effect: Appearance
  of Neutral Modes},'' {\em {Phys. Rev. Lett.}}, vol.~122, p.~246801, Jun 2019.

\bibitem{Law2006}
K.~T. Law, D.~E. Feldman, and Y.~Gefen, ``{Electronic Mach-Zehnder
  interferometer as a tool to probe fractional statistics},'' {\em {Phys. Rev.
  B}}, vol.~74, p.~045319, Jul 2006.

\bibitem{Feldman2007}
D.~E. Feldman, Y.~Gefen, A.~Kitaev, K.~T. Law, and A.~Stern, ``{Shot noise in
  an anyonic Mach-Zehnder interferometer},'' {\em {Phys. Rev. B}}, vol.~76,
  p.~085333, Aug 2007.

\bibitem{Byers1961}
N.~Byers and C.~N. Yang, ``{Theoretical Considerations Concerning Quantized
  Magnetic Flux in Superconducting Cylinders},'' {\em {Phys. Rev. Lett.}},
  vol.~7, pp.~46--49, Jul 1961.

\bibitem{Feldman2006}
D.~E. Feldman and A.~Kitaev, ``{Detecting Non-Abelian Statistics with an
  Electronic Mach-Zehnder Interferometer},'' {\em {Phys. Rev. Lett.}}, vol.~97,
  p.~186803, Nov 2006.

\bibitem{Jonckheere2005}
T.~Jonckheere, P.~Devillard, A.~Cr\'epieux, and T.~Martin, ``{Electronic
  Mach-Zehnder interferometer in the fractional quantum Hall effect},'' {\em
  Phys. Rev. B}, vol.~72, p.~201305, Nov 2005.

\bibitem{Kane1994}
C.~L. Kane, M.~P.~A. Fisher, and J.~Polchinski, ``{Randomness at the edge:
  Theory of quantum Hall transport at filling \ensuremath{\nu}=2/3},'' {\em
  Phys. Rev. Lett.}, vol.~72, pp.~4129--4132, Jun 1994.

\bibitem{Kane1995}
C.~L. Kane and M.~P.~A. Fisher, ``{Impurity scattering and transport of
  fractional quantum Hall edge states},'' {\em Phys. Rev. B}, vol.~51,
  pp.~13449--13466, May 1995.

\bibitem{MacDonald1990}
A.~H. MacDonald, ``{Edge states in the fractional-quantum-Hall-effect
  regime},'' {\em Phys. Rev. Lett.}, vol.~64, pp.~220--223, Jan 1990.

\bibitem{Wen1990}
X.~G. Wen, ``{Electrodynamical properties of gapless edge excitations in the
  fractional quantum Hall states},'' {\em Phys. Rev. Lett.}, vol.~64,
  pp.~2206--2209, Apr 1990.

\bibitem{Wen1990a}
X.~G. Wen, ``{Chiral Luttinger liquid and the edge excitations in the
  fractional quantum Hall states},'' {\em Phys. Rev. B}, vol.~41,
  pp.~12838--12844, Jun 1990.

\bibitem{Goldstein2016}
M.~Goldstein and Y.~Gefen, ``{Suppression of Interference in Quantum Hall
  Mach-Zehnder Geometry by Upstream Neutral Modes},'' {\em {Phys. Rev. Lett.}},
  vol.~117, p.~276804, Dec 2016.

\bibitem{Wang2013}
J.~Wang, Y.~Meir, and Y.~Gefen, ``{Edge Reconstruction in the $\nu=2/3$
  Fractional Quantum Hall State},'' {\em Phys. Rev. Lett.}, vol.~111,
  p.~246803, Dec 2013.

\bibitem{Sabo2017}
R.~Sabo, I.~Gurman, A.~Rosenblatt, F.~Lafont, D.~Banitt, J.~Park, M.~Heiblum,
  Y.~Gefen, V.~Umansky, and D.~Mahalu, ``{Edge reconstruction in fractional
  quantum Hall states},'' {\em Nat. Phys.}, vol.~13, p.~491, 2017.

\bibitem{Park2021}
J.~Park, B.~Rosenow, and Y.~Gefen, ``{Symmetry-related transport on a
  fractional quantum Hall edge},'' 2021.

\bibitem{Chamon1997}
C.~de~C.~Chamon, D.~E. Freed, S.~A. Kivelson, S.~L. Sondhi, and X.~G. Wen,
  ``{Two point-contact interferometer for quantum Hall systems},'' {\em {Phys.
  Rev. B}}, vol.~55, pp.~2331--2343, Jan 1997.

\bibitem{Halperin2011}
B.~I. Halperin, A.~Stern, I.~Neder, and B.~Rosenow, ``{Theory of the
  Fabry-P\'erot quantum Hall interferometer},'' {\em {Phys. Rev. B}}, vol.~83,
  p.~155440, Apr 2011.

\bibitem{Jain1989}
J.~K. Jain, ``{Composite-fermion approach for the fractional quantum Hall
  effect},'' {\em {Phys. Rev. Lett.}}, vol.~63, pp.~199--202, Jul 1989.

\bibitem{Inoue2014}
H.~Inoue, A.~Grivnin, Y.~Ronen, M.~Heiblum, V.~Umansky, and D.~Mahalu,
  ``{Proliferation of neutral modes in fractional quantum Hall states},'' {\em
  {Nat. Commun.}}, vol.~5, p.~4067, 2014.

\bibitem{Grivnin2014}
A.~Grivnin, H.~Inoue, Y.~Ronen, Y.~Baum, M.~Heiblum, V.~Umansky, and D.~Mahalu,
  ``{Nonequilibrated Counterpropagating Edge Modes in the Fractional Quantum
  Hall Regime},'' {\em {Phys. Rev. Lett.}}, vol.~113, p.~266803, Dec 2014.

\bibitem{Rosenow2007}
B.~Rosenow and B.~I. Halperin, ``{Influence of Interactions on Flux and
  Back-Gate Period of Quantum Hall Interferometers},'' {\em {Phys. Rev.
  Lett.}}, vol.~98, p.~106801, Mar 2007.

\bibitem{NgoDinh2012}
S.~Ngo~Dinh and D.~A. Bagrets, ``{Influence of Coulomb interaction on the
  Aharonov-Bohm effect in an electronic Fabry-P\'erot interferometer},'' {\em
  {Phys. Rev. B}}, vol.~85, p.~073403, Feb 2012.

\bibitem{Wees1989}
B.~J. van Wees, L.~P. Kouwenhoven, C.~J. P.~M. Harmans, J.~G. Williamson, C.~E.
  Timmering, M.~E.~I. Broekaart, C.~T. Foxon, and J.~J. Harris, ``Observation
  of zero-dimensional states in a one-dimensional electron interferometer,''
  {\em {Phys. Rev. Lett.}}, vol.~62, pp.~2523--2526, May 1989.

\bibitem{Camino2005}
F.~E. Camino, W.~Zhou, and V.~J. Goldman, ``{Aharonov-Bohm Superperiod in a
  Laughlin Quasiparticle Interferometer},'' {\em {Phys. Rev. Lett.}}, vol.~95,
  p.~246802, Dec 2005.

\bibitem{Camino2007}
F.~E. Camino, W.~Zhou, and V.~J. Goldman, ``{$e/3$ Laughlin Quasiparticle
  Primary-Filling $\ensuremath{\nu}=1/3$ Interferometer},'' {\em {Phys. Rev.
  Lett.}}, vol.~98, p.~076805, Feb 2007.

\bibitem{Godfrey2007}
M.~D. Godfrey, P.~Jiang, W.~Kang, S.~H. Simon, K.~W. Baldwin, L.~N. Pfeiffer,
  and K.~W. West, ``{Aharonov-Bohm-Like Oscillations in Quantum Hall
  Corrals},'' 2007.

\bibitem{Deviatov2008}
E.~V. Deviatov and A.~Lorke, ``{Experimental realization of a Fabry-Perot-type
  interferometer by copropagating edge states in the quantum Hall regime},''
  {\em {Phys. Rev. B}}, vol.~77, p.~161302, Apr 2008.

\bibitem{Zhang2009}
Y.~Zhang, D.~T. McClure, E.~M. Levenson-Falk, C.~M. Marcus, L.~N. Pfeiffer, and
  K.~W. West, ``{Distinct signatures for Coulomb blockade and Aharonov-Bohm
  interference in electronic Fabry-Perot interferometers},'' {\em {Phys. Rev.
  B}}, vol.~79, p.~241304, Jun 2009.

\bibitem{Willett2009}
R.~L. Willett, L.~N. Pfeiffer, and K.~W. West, ``Measurement of filling factor
  5/2 quasiparticle interference with observation of charge e/4 and e/2 period
  oscillations,'' {\em Proc. Natl. Acad. Sci.}, vol.~106, no.~22,
  pp.~8853--8858, 2009.

\bibitem{Lin2009}
P.~V. Lin, F.~E. Camino, and V.~J. Goldman, ``{Electron interferometry in the
  quantum Hall regime: Aharonov-Bohm effect of interacting electrons},'' {\em
  {Phys. Rev. B}}, vol.~80, p.~125310, Sep 2009.

\bibitem{Lin2009a}
P.~V. Lin, F.~E. Camino, and V.~J. Goldman, ``{Superperiods in interference of
  $e/3$ Laughlin quasiparticles encircling filling 2/5 fractional quantum Hall
  island},'' {\em {Phys. Rev. B}}, vol.~80, p.~235301, Dec 2009.

\bibitem{McClure2009}
D.~T. McClure, Y.~Zhang, B.~Rosenow, E.~M. Levenson-Falk, C.~M. Marcus, L.~N.
  Pfeiffer, and K.~W. West, ``{Edge-State Velocity and Coherence in a Quantum
  Hall Fabry-P\'erot Interferometer},'' {\em {Phys. Rev. Lett.}}, vol.~103,
  p.~206806, Nov 2009.

\bibitem{Ofek2010}
N.~Ofek, A.~Bid, M.~Heiblum, A.~Stern, V.~Umansky, and D.~Mahalu, ``{Role of
  interactions in an electronic Fabry{\textendash}Perot interferometer
  operating in the quantum Hall effect regime},'' {\em Proc. Natl. Acad. Sci.},
  vol.~107, no.~12, pp.~5276--5281, 2010.

\bibitem{McClure2012}
D.~T. McClure, W.~Chang, C.~M. Marcus, L.~N. Pfeiffer, and K.~W. West,
  ``{Fabry-Perot Interferometry with Fractional Charges},'' {\em {Phys. Rev.
  Lett.}}, vol.~108, p.~256804, Jun 2012.

\bibitem{Choi2015}
H.~K. Choi, I.~Sivan, A.~Rosenblatt, M.~Heiblum, V.~Umansky, and D.~Mahalu,
  ``{Robust electron pairing in the integer quantum Hall effect regime},'' {\em
  Nat. Commun.}, vol.~6, p.~7435, jun 2015.

\bibitem{Sivan2016}
I.~Sivan, H.~K. Choi, J.~Park, A.~Rosenblatt, Y.~Gefen, D.~Mahalu, and
  V.~Umansky, ``{Observation of interaction-induced modulations of a quantum
  Hall liquid's area},'' {\em Nat. Commun.}, vol.~7, p.~12184, jul 2016.

\bibitem{Sivan2018}
I.~Sivan, R.~Bhattacharyya, H.~K. Choi, M.~Heiblum, D.~E. Feldman, D.~Mahalu,
  and V.~Umansky, ``{Interaction-induced interference in the integer quantum
  Hall effect},'' {\em {Phys. Rev. B}}, vol.~97, p.~125405, Mar 2018.

\bibitem{Nakamura2019}
J.~Nakamura, S.~Fallahi, H.~Sahasrabudhe, R.~Rahman, S.~Liang, G.~C. Gardner,
  and M.~J. Manfra, ``{Aharonov{\textendash}Bohm interference of fractional
  quantum Hall edge modes},'' {\em Nat. Phys.}, vol.~15, pp.~563--569, mar
  2019.

\bibitem{Roosli2020}
M.~P. R\"o\"osli, L.~Brem, B.~Kratochwil, G.~Nicol\'{\i}, B.~A. Braem,
  S.~Hennel, P.~M\"arki, M.~Berl, C.~Reichl, W.~Wegscheider, K.~Ensslin,
  T.~Ihn, and B.~Rosenow, ``{Observation of quantum Hall interferometer phase
  jumps due to a change in the number of bulk quasiparticles},'' {\em Phys.
  Rev. B}, vol.~101, p.~125302, Mar 2020.

\bibitem{Simmons1989}
J.~A. Simmons, H.~P. Wei, L.~W. Engel, D.~C. Tsui, and M.~Shayegan,
  ``{Resistance fluctuations in narrow AlGaAs/GaAs heterostructures: Direct
  evidence of fractional charge in the fractional quantum Hall effect},'' {\em
  Phys. Rev. Lett.}, vol.~63, pp.~1731--1734, Oct 1989.

\bibitem{Goldman1995}
V.~J. Goldman and B.~Su, ``{Resonant Tunneling in the Quantum Hall Regime:
  Measurement of Fractional Charge},'' {\em Science}, vol.~267, no.~5200,
  pp.~1010--1012, 1995.

\bibitem{Maasilta2000}
I.~J. Maasilta and V.~J. Goldman, ``{Tunneling through a Coherent ``Quantum
  Antidot Molecule''},'' {\em Phys. Rev. Lett.}, vol.~84, pp.~1776--1779, Feb
  2000.

\bibitem{Goldman2001}
V.~J. Goldman, I.~Karakurt, J.~Liu, and A.~Zaslavsky, ``{Invariance of charge
  of Laughlin quasiparticles},'' {\em Phys. Rev. B}, vol.~64, p.~085319, Aug
  2001.

\bibitem{Goldman2003}
V.~J. Goldman, ``The quantum antidot electrometer: direct observation of
  fractional charge,'' {\em Journal of the Korean Physical Society}, vol.~39,
  no.~3, pp.~512--518, 2003.

\bibitem{Kivelson1990}
S.~Kivelson, ``Semiclassical theory of localized many-anyon states,'' {\em
  Phys. Rev. Lett.}, vol.~65, pp.~3369--3372, Dec 1990.

\bibitem{Lee1990}
P.~A. Lee, ``{Comment on ``Resistance fluctuations in narrow AlGaAs/GaAs
  heterostructures: Direct evidence of fractional charge in the fractional
  quantum Hall effect''},'' {\em Phys. Rev. Lett.}, vol.~65, pp.~2206--2206,
  Oct 1990.

\bibitem{Thouless1991}
D.~J. Thouless and Y.~Gefen, ``{Fractional quantum Hall effect and multiple
  Aharonov-Bohm periods},'' {\em Phys. Rev. Lett.}, vol.~66, pp.~806--809, Feb
  1991.

\bibitem{Gefen1993}
Y.~Gefen and D.~J. Thouless, ``{Detection of fractional charge and quenching of
  the quantum Hall effect},'' {\em Phys. Rev. B}, vol.~47, pp.~10423--10436,
  Apr 1993.

\bibitem{Camino2008}
F.~E. Camino, W.~Zhou, and V.~J. Goldman, ``{Experimental realization of
  Laughlin quasiparticle interferometers},'' {\em Physica E}, vol.~40, no.~5,
  pp.~949--953, 2008.

\bibitem{Ferraro2017}
D.~Ferraro and E.~Sukhorukov, ``{Interaction effects in a multi-channel
  Fabry-P{\'e}rot interferometer in the Aharonov-Bohm regime},'' {\em SciPost
  Phys.}, vol.~3, p.~014, 2017.

\bibitem{Frigeri2019}
G.~A. Frigeri, D.~D. Scherer, and B.~Rosenow, ``{Sub-periods and apparent
  pairing in integer quantum Hall interferometers},'' {\em {EPL} (Europhysics
  Letters)}, vol.~126, p.~67007, jul 2019.

\bibitem{Frigeri2020}
G.~A. Frigeri and B.~Rosenow, ``{Electron pairing in the quantum Hall regime
  due to neutralon exchange},'' {\em Phys. Rev. Research}, vol.~2, p.~043396,
  Dec 2020.

\bibitem{Manfra2014}
M.~J. Manfra, ``{Molecular Beam Epitaxy of Ultra-High-Quality AlGaAs/GaAs
  Heterostructures: Enabling Physics in Low-Dimensional Electronic Systems},''
  {\em Annu. Rev. Condens. Matter Phys.}, vol.~5, no.~1, pp.~347--373, 2014.

\bibitem{Sahasrabudhe2018}
H.~Sahasrabudhe, B.~Novakovic, J.~Nakamura, S.~Fallahi, M.~Povolotskyi,
  G.~Klimeck, R.~Rahman, and M.~J. Manfra, ``{Optimization of edge state
  velocity in the integer quantum Hall regime},'' {\em {Phys. Rev. B}},
  vol.~97, p.~085302, Feb 2018.

\bibitem{Rosenow2020}
B.~Rosenow and A.~Stern, ``{Flux Superperiods and Periodicity Transitions in
  Quantum Hall Interferometers},'' {\em {Phys. Rev. Lett.}}, vol.~124,
  p.~106805, Mar 2020.

\bibitem{DasSarma2005}
S.~Das~Sarma, M.~Freedman, and C.~Nayak, ``{Topologically Protected Qubits from
  a Possible Non-Abelian Fractional Quantum Hall State},'' {\em Phys. Rev.
  Lett.}, vol.~94, p.~166802, Apr 2005.

\bibitem{Fendley2005}
P.~Fendley and E.~Fradkin, ``{Realizing non-Abelian statistics in
  time-reversal-invariant systems},'' {\em Phys. Rev. B}, vol.~72, p.~024412,
  Jul 2005.

\bibitem{Freedman2006}
M.~Freedman, C.~Nayak, and K.~Walker, ``{Towards universal topological quantum
  computation in the $\ensuremath{\nu}=\frac{5}{2}$ fractional quantum Hall
  state},'' {\em Phys. Rev. B}, vol.~73, p.~245307, Jun 2006.

\bibitem{Kitaev2006}
A.~Kitaev, ``Anyons in an exactly solved model and beyond,'' {\em Ann.
  Physics}, vol.~321, no.~1, pp.~2--111, 2006.

\bibitem{Moore1991}
G.~Moore and N.~Read, ``{Nonabelions in the fractional quantum Hall effect},''
  {\em Nuclear Physics B}, vol.~360, no.~2, pp.~362 -- 396, 1991.

\bibitem{Morf1998}
R.~H. Morf, ``{Transition from Quantum Hall to Compressible States in the
  Second Landau Level: New Light on the $\nu=5/2$ Enigma},'' {\em {Phys. Rev.
  Lett.}}, vol.~80, pp.~1505--1508, Feb 1998.

\bibitem{Chung2003}
Y.~C. Chung, M.~Heiblum, and V.~Umansky, ``Scattering of bunched fractionally
  charged quasiparticles,'' {\em Phys. Rev. Lett.}, vol.~91, p.~216804, Nov
  2003.

\bibitem{Levin2007}
M.~Levin, B.~I. Halperin, and B.~Rosenow, ``{Particle-Hole Symmetry and the
  Pfaffian State},'' {\em Phys. Rev. Lett.}, vol.~99, p.~236806, Dec 2007.

\bibitem{Bishara2008}
W.~Bishara, G.~A. Fiete, and C.~Nayak, ``{Quantum Hall states at
  $\ensuremath{\nu}=\frac{2}{k+2}$: Analysis of the particle-hole conjugates of
  the general level-$k$ Read-Rezayi states},'' {\em Phys. Rev. B}, vol.~77,
  p.~241306, Jun 2008.

\bibitem{Dolev2008}
M.~Dolev, M.~Heiblum, V.~Umansky, A.~Stern, and D.~Mahalu, ``{Observation of a
  quarter of an electron charge at the $\nu = 5/2$ quantum Hall state},'' {\em
  Nature}, vol.~452, p.~829, 2008.

\bibitem{Carrega2011}
M.~Carrega, D.~Ferraro, A.~Braggio, N.~Magnoli, and M.~Sassetti, ``{Anomalous
  Charge Tunneling in Fractional Quantum Hall Edge States at a Filling Factor
  $\ensuremath{\nu}=5/2$},'' {\em Phys. Rev. Lett.}, vol.~107, p.~146404, Sep
  2011.

\bibitem{Stern2006}
A.~Stern and B.~I. Halperin, ``{Proposed Experiments to Probe the Non-Abelian
  $\ensuremath{\nu}=5/2$ Quantum Hall State},'' {\em Phys. Rev. Lett.},
  vol.~96, p.~016802, Jan 2006.

\bibitem{Fendley2007}
P.~Fendley, M.~P.~A. Fisher, and C.~Nayak, ``{Edge states and tunneling of
  non-Abelian quasiparticles in the $\ensuremath{\nu}=5/2$ quantum Hall state
  and $p+ip$ superconductors},'' {\em Phys. Rev. B}, vol.~75, p.~045317, Jan
  2007.

\bibitem{Bishara2009}
W.~Bishara, P.~Bonderson, C.~Nayak, K.~Shtengel, and J.~K. Slingerland,
  ``{Interferometric signature of non-Abelian anyons},'' {\em {Phys. Rev. B}},
  vol.~80, p.~155303, Oct 2009.

\bibitem{Tiemann2012}
L.~Tiemann, G.~Gamez, N.~Kumada, and K.~Muraki, ``{Unraveling the Spin
  Polarization of the $\nu$ = 5/2 Fractional Quantum Hall State},'' {\em
  {Science}}, vol.~335, no.~6070, pp.~828--831, 2012.

\bibitem{Banerjee2018}
M.~Banerjee, M.~Heiblum, V.~Umansky, D.~E. Feldman, Y.~Oreg, and A.~Stern,
  ``{Observation of half-integer thermal Hall conductance},'' {\em Nature},
  vol.~559, no.~7713, pp.~205--210, 2018.

\bibitem{Willett2019}
R.~L. Willett, K.~Shtengel, C.~Nayak, L.~N. Pfeiffer, Y.~J. Chung, M.~L.
  Peabody, K.~W. Baldwin, and K.~W. West, ``{Interference measurements of
  non-Abelian e/4 \& Abelian e/2 quasiparticle braiding},'' 2019.

\bibitem{Willett2013}
R.~L. Willett, C.~Nayak, K.~Shtengel, L.~N. Pfeiffer, and K.~W. West,
  ``{Magnetic-Field-Tuned Aharonov-Bohm Oscillations and Evidence for
  Non-Abelian Anyons at $\ensuremath{\nu}=5/2$},'' {\em {Phys. Rev. Lett.}},
  vol.~111, p.~186401, Oct 2013.

\bibitem{Deprez2021}
C.~D{\'e}prez, L.~Veyrat, H.~Vignaud, G.~Nayak, K.~Watanabe, T.~Taniguchi,
  F.~Gay, H.~Sellier, and B.~Sac{\'e}p{\'e}, ``{A tunable Fabry--P{\'e}rot
  quantum Hall interferometer in graphene},'' {\em Nature Nanotechnology},
  2021.

\bibitem{Ronen2021}
Y.~Ronen, T.~Werkmeister, D.~Haie~Najafabadi, A.~T. Pierce, L.~E. Anderson,
  Y.~J. Shin, S.~Y. Lee, Y.~H. Lee, B.~Johnson, K.~Watanabe, T.~Taniguchi,
  A.~Yacoby, and P.~Kim, ``{Aharonov--Bohm effect in graphene-based
  Fabry--P{\'e}rot quantum Hall interferometers},'' {\em Nature
  Nanotechnology}, 2021.

\bibitem{Ferraro2017a}
D.~Ferraro, T.~Jonckheere, J.~Rech, and T.~Martin, ``{Electronic quantum optics
  beyond the integer quantum Hall effect},'' {\em physica status solidi (b)},
  vol.~254, no.~3, p.~1600531, 2017.

\bibitem{Oliver1999}
W.~D. Oliver, J.~Kim, R.~C. Liu, and Y.~Yamamoto, ``{Hanbury Brown and
  Twiss-Type Experiment with Electrons},'' {\em Science}, vol.~284, no.~5412,
  pp.~299--301, 1999.

\bibitem{Henny1999}
M.~Henny, S.~Oberholzer, C.~Strunk, T.~Heinzel, K.~Ensslin, M.~Holland, and
  C.~Sch{\"o}nen\-berger, ``{The Fermionic Hanbury Brown and Twiss
  Experiment},'' {\em Science}, vol.~284, no.~5412, pp.~296--298, 1999.

\bibitem{Oberholzer2000}
S.~Oberholzer, M.~Henny, C.~Strunk, C.~Sch{\"o}nenberger, T.~Heinzel,
  K.~Ensslin, and M.~Holland, ``{The Hanbury Brown and Twiss experiment with
  fermions},'' {\em Physica E: Low-dimensional Systems and Nanostructures},
  vol.~6, no.~1, pp.~314--317, 2000.

\bibitem{Safi2001}
I.~Safi, P.~Devillard, and T.~Martin, ``{Partition Noise and Statistics in the
  Fractional Quantum Hall Effect},'' {\em Phys. Rev. Lett.}, vol.~86,
  pp.~4628--4631, May 2001.

\bibitem{Samuelsson2004}
P.~Samuelsson, E.~V. Sukhorukov, and M.~B\"uttiker, ``{Two-Particle
  Aharonov-Bohm Effect and Entanglement in the Electronic Hanbury Brown--Twiss
  Setup},'' {\em Phys. Rev. Lett.}, vol.~92, p.~026805, Jan 2004.

\bibitem{Campagnano2012}
G.~Campagnano, O.~Zilberberg, I.~V. Gornyi, D.~E. Feldman, A.~C. Potter, and
  Y.~Gefen, ``{Hanbury Brown--Twiss Interference of Anyons},'' {\em Phys. Rev.
  Lett.}, vol.~109, p.~106802, Sep 2012.

\bibitem{Campagnano2013}
G.~Campagnano, O.~Zilberberg, I.~V. Gornyi, and Y.~Gefen, ``{Hanbury Brown and
  Twiss correlations in quantum Hall systems},'' {\em Phys. Rev. B}, vol.~88,
  p.~235415, Dec 2013.

\bibitem{Jonckheere2012}
T.~Jonckheere, J.~Rech, C.~Wahl, and T.~Martin, ``{Electron and hole
  Hong-Ou-Mandel interferometry},'' {\em Phys. Rev. B}, vol.~86, p.~125425, Sep
  2012.

\bibitem{Freulon2015}
V.~Freulon, A.~Marguerite, J.-M. Berroir, B.~Pla{\c{c}}ais, A.~Cavanna, Y.~Jin,
  and G.~F{\`{e}}ve, ``{Hong-Ou-Mandel experiment for temporal investigation of
  single-electron fractionalization},'' {\em Nat. Commun.}, vol.~6, p.~6854,
  apr 2015.

\bibitem{Beenakker}
C.~W.~J. Beenakker and H.~van Houten, ``{Quantum Transport in Semiconductor
  Nanostructures},'' {\em Solid State Phys}, vol.~44, pp.~1--228, 1991.

\bibitem{Landauer}
R.~Landauer, ``{Spatial Variation of Currents and Fields Due to Localized
  Scatterers in Metallic Conduction},'' {\em IBM J. Res. Dev.}, vol.~1, p.~223,
  1957.

\bibitem{Landauer2}
R.~Landauer, ``{Spatial variation of currents and fields due to localized
  scatterers in metallic conduction},'' {\em IBM J. Res. Dev.}, vol.~32,
  p.~306, 1988.

\bibitem{Conductance1}
B.~J. van Wees, H.~van Houten, C.~W.~J. Beenakker, J.~G. Willamson, L.~P.
  Kouwenhoven, D.~van~der Marel, and C.~T. Foxon, ``{Quantized Conductance of
  Point Contacts in a Two-Dimensional Electron Gas},'' {\em Phys. Rev. Lett.},
  vol.~60, p.~848, 1988.

\bibitem{Conductance2}
D.~A. Wharam, T.~J. Thornton, R.~Newbury, M.~Pepper, H.~Ahmed, J.~E.~F. Frost,
  D.~G. Hasko, D.~C. Peacock, D.~A. Ritchie, and G.~A.~C. Jones,
  ``One-dimensional transport and the quantisation of the ballistic
  resistance,'' {\em J. Phys. C}, vol.~21, p.~L209, 1988.

\bibitem{Schapers}
T.~Sch\"apers, {\em Superconductor/semiconductor junctions}.
\newblock Springer, 2001.

\bibitem{Braggio2012}
A.~Braggio, D.~Ferraro, M.~Carrega, N.~Magnoli, and M.~Sassetti,
  ``{Environmental induced renormalization effects in quantum Hall edge states
  due to 1/f noise and dissipation},'' {\em New Journal of Physics}, vol.~14,
  p.~093032, sep 2012.

\bibitem{Cohen2019}
Y.~Cohen, Y.~Ronen, W.~Yang, D.~Banitt, J.~Park, M.~Heiblum, A.~D. Mirlin,
  Y.~Gefen, and V.~Umansky, ``{Synthesizing a $\nu=2/3$ fractional quantum Hall
  effect edge state from counter-propagating $\nu=1$ and $\nu=1/3$ states},''
  {\em Nature Communications}, vol.~10, no.~1, p.~1920, 2019.

\bibitem{Nosiglia2018}
C.~Nosiglia, J.~Park, B.~Rosenow, and Y.~Gefen, ``{Incoherent transport on the
  $\nu=2/3$ quantum Hall edge},'' {\em Phys. Rev. B}, vol.~98, p.~115408, Sep
  2018.

\bibitem{Protopopov2017}
I.~Protopopov, Y.~Gefen, and A.~Mirlin, ``{Transport in a disordered $\nu =
  2/3$ fractional quantum Hall junction},'' {\em Annals of Physics}, vol.~385,
  pp.~287--327, 2017.

\end{thebibliography}

\end{document}